\documentclass[11pt,preprint]{aastex}
\usepackage{epsf,epsfig}
\def\avg#1{\langle #1 \rangle}

\def\Sth{\mbox{${\cal S}_{\rm th}$}}
\def\etal{{\it et al.~}}
\def\iso#1#2{\mbox{${}^{#2}{\rm #1}$}}
\def\he#1{\iso{He}{#1}}
\def\be#1{\iso{Be}{#1}}
\def\li#1{\iso{Li}{#1}}
\def\b1#1{\iso{B}{1#1}}

\def\omegab{\Omega_{\rm B}}

\def\ee#1#2{#1 \times 10^{#2}}

\def\pref#1{(\ref{#1})}
\def\pcite#1{\cite{#1}}

\def\beq{\begin{equation}}
\def\eeq{\end{equation}}
\def\beqar{\begin{eqnarray}}
\def\eeqar{\end{eqnarray}}

\begin{document}

\title{THE NACRE THERMONUCLEAR REACTION COMPILATION \\
AND BIG BANG NUCLEOSYNTHESIS}
 
\author{Richard H. Cyburt}
\affil{Department of Physics, University of Illinois,
Urbana, IL 61801, USA}

\author{Brian D. Fields}
\affil{Center for Theoretical Astrophysics, Department of Astronomy,
University of Illinois, Urbana, IL 61801, USA}

\author{Keith~A.~Olive}
\affil{TH Division, CERN, Geneva, Switzerland \\
also Theoretical Physics Institute, School of Physics and Astronomy \\
University of Minnesota, Minneapolis, MN 55455 USA }

\begin{abstract}
The theoretical predictions of big bang nucleosynthesis (BBN)
are dominated by uncertainties in the input
nuclear reaction cross sections.
In this paper, 
we examine the impact on BBN of the
recent compilation of nuclear data and thermonuclear
reactions rates 
by the NACRE collaboration.  
We confirm that the adopted rates do not make
large overall changes in central values of predictions,
but do affect the magnitude of the uncertainties in these
predictions.
Therefore, we then examine in detail the uncertainties
in the individual reaction rates considered by NACRE.
When the error estimates by NACRE are treated
as $1\sigma$ limits, the resulting
BBN error budget
is similar to those of previous tabulations.
We propose two new procedures for deriving reaction rate
uncertainties from the nuclear data: one which sets lower 
limits to the error, and one which we believe is a reasonable
description of the present error budget.
We propagate these uncertainty estimates through the BBN
code, and find that when
the nuclear data errors are described most accurately,
the resulting light element uncertainties are notably smaller
than in some previous tabulations, but larger than
others.
Using these results, we derive limits on the
cosmic baryon-to-photon ratio $\eta$, and compare
this to independent limits on $\eta$ from recent 
balloon-borne measurements of the cosmic microwave background radiation
(CMB).
We discuss means to improve the BBN results
via key nuclear reaction
measurements and light element observations.

\end{abstract}


\newpage

\section{Introduction}

Big bang nucleosynthesis (BBN) has long played a
key role in the standard cosmology.
The concordance between the predicted and observed light element
abundances (Walker \etal 1991; Olive, Steigman, \& Walker 2000) gives
confidence that the basic framework is working down to epochs of $t
\sim 1$ sec, and $z \sim 10^{10}$. The theory-observation comparison
also  probes particle physics in
the early universe (Sarkar 1996; Olive \& Thomas 1999; Lisi, Sarkar, \& Villante 1999),
and  quantifies the
allowed range in the baryon-to-photon ratio 
$\eta = n_{\rm B}/n_\gamma$ (Fields \& Olive 1996, Fields \etal 1996, Schramm \&
Turner 1998) and thus the cosmic baryon density parameter 
$\Omega_{\rm B} h^2$ (where
$\Omega_i = \rho_i/\rho_{\rm crit} = 8\pi G \rho_i/3H_0^2$, and
$H_0 = 100 h \ {\rm km \ s^{-1} \ Mpc^{-1}}$).

Over the past decade, a major thrust of research in BBN has been
towards increasing the rigor of the analysis.
On the theory side, the key 
innovation was to calculate the errors in the light element
predictions in a systematic and statistically careful way.  
This was done using Monte Carlo analyses 
(Krauss \& Romanelli 1990; Smith, Kawano, \& Malaney 1993; Kernan \& Krauss 1995; 
Hata et al.\ 1995; Fiorentini \etal 1998; 
Nollett \& Burles 2000),
which account for nuclear reaction uncertainties and
their propagation into uncertainties in the light element
abundance predictions.  
These calculations are essential because they allow
for a careful statistical comparison of BBN theory
with observational constraints; in addition, they point
the way toward improvements in the theory calculation.
Just as BBN theoretical and observational ingredients have
been sharpened, so have the tools for comparing the two.
Statistical tests involving likelihood analyses
(Hata et al.\ 1995; Fields, Kainulainen, Olive, \& Thomas 1996;
Lisi, Sarkar, \& Villante 1999)
have been developed and performed.

A focus on rigor in BBN will soon be rewarded,
as cosmology moves toward a precision era.
Specifically, the observations of the
anisotropies in the cosmic microwave background (CMB)
will allow for very precise determination of
cosmological parameters,
including $\Omega_{\rm B} h^2$ (or, equivalently, $\eta$).
A comparison of $\eta$ as determined by BBN and the
CMB will provide a fundamental test of cosmology.
In addition, as pointed out by Schramm \& Turner (1998) and
as shown in detail below,
increasingly accurate CMB data
can ultimately transform BBN into a much sharper probe of the
early universe and of chemical evolution.

Along with the neutron lifetime, eleven key
nuclear reaction rates represent the
dominant sources of error the BBN calculation
(Smith, Kawano, \& Malaney \cite{skm}, hereafter SKM;
and \S \ref{sect:bbn_impact} below). Thus, the choice of
thermonuclear reaction rates, and their uncertainties,
will determine the accuracy of the final predictions.
Most recent work on BBN has used the Caughlan \& Fowler \pcite{cf}
compilation,
with updates due to SKM and others.
The SKM error budget has been the standard
for all of the subsequent Monte Carlo work 
except for that of 
Nollett \& Burles \pcite{nb},
who create their own rates
from the nuclear data, but do not present the
thermonuclear rates by themselves.
The recent work by NACRE (Angulo et al.\ \cite{nacre})\footnote{
{\tt \url{http://pntpm.ulb.ac.be/nacre.htm}}
}
represents a significant new effort to
critically evaluate nuclear cross section data and
derive thermonuclear rates.  Moreover, the
NACRE compilation includes estimates
of uncertainties in the reaction rates.  

We therefore have examined the impact of the NACRE rates on BBN.
This issue
has also been studied by Vangioni-Flam, Coc, \& Cass\'{e} \cite{vcc}.
These authors considered the impact of the error range estimated by
NACRE by calculating the effect
of {\em individual} rates on the primordial abundances.
Vangioni-Flam, Coc, \& Cass\'{e} \cite{vcc} then
estimated the effect of
the {\em ensemble} of rates by placing all rates
at their maximum and
minimum variation.
Here, we will extend this work in several ways.
We will use a full Monte-Carlo treatment, which allows for
a quantitative statement about the propagation of the errors.
In addition, we examine the impact of different 
error assignments:  (1) those of the NACRE group themselves,
which give useful estimates of the range of uncertainty but
are not defined in a uniform manner; as well as 
(2) the results of our own error analyses, which
we derive from the nuclear data according to simple
but uniform procedures.

This paper is organized as follows.
We will describe 
the NACRE recommended rate
uncertainties in \S \ref{sect:nacre}.
In \S \ref{sect:errors}
we compare the NACRE uncertainty estimations
with other estimates based on the nuclear reaction data.
These different error estimates are
used to predict light element uncertainties via Monte Carlo  BBN 
calculations (\S \ref{sect:bbn_impact}). The results
are compared with observations via likelihood
analyses in \S \ref{sect:like}, and 
discussed in terms of the predictions 
for $\eta$. We discuss the agreement of
our predictions with current CMB data in \S \ref{sect:cmb},
and anticipate the impact of high-quality data which
will arise from future CMB space-based experiments.
Our conclusions are summarized in \S \ref{sect:conclude}.

\section{The NACRE Compilation}
\label{sect:nacre}

In an effort to update the thermonuclear reaction rate compilation by
Caughlan \& Fowler (1988), the NACRE collaboration has presented a
detailed analysis of 86 charge induced nuclear reactions.  Out of
these reactions we will discuss the 7 reactions NACRE has in common
with the 11 reactions from SKM\pcite{skm}, whose errors dominate the
uncertainties in the abundances.(see \S 2.3)

\subsection{Reaction Rate Formalism}

The nuclear reaction inputs to BBN take the form of thermal rates.
These rates are computed by averaging nuclear reaction cross sections over a
Maxwell-Boltzmann distribution of energies.  The thermonuclear
reaction rate at some temperature $T$, is given by the following integral:
\beq
\label{eq:thermrate}
N_A\langle \sigma v \rangle = N_A\left( \frac{8}{\pi \mu (k_{B}T)^3}
\right)^{\frac{1}{2}} \int_{0}^{\infty} \sigma(E)E \exp \left(-
\frac{E}{k_{B}T} \right) dE
\eeq
where $N_A$ is Avogadro's number, $v$ the relative velocity, $\mu$ gives
the reduced mass of the nuclei, $k_{B}$ Boltzmann's constant and
$\sigma(E)$ yields the cross section at center of mass energy $E$.

  The charge induced cross sections can be decomposed into
\beq
\sigma(E) = \frac{S(E)\exp \left( -2\pi\zeta \right)}{E},
\eeq
where $S(E)$ gives the astrophysical $S$-factor, $\zeta$ is the Sommerfeld
parameter, defined by
\beq
\zeta = Z_{1}Z_{2}\left( \frac{e^2}{\hbar c} \right)\left(
\frac{c}{v} \right) = Z_{1}Z_{2}\left( \frac{e^2}{\hbar c}
\right)\left( \frac{\mu c^2}{2E} \right)^{\frac{1}{2}} .
\eeq
Here the $Z_{i}$'s are the charges of the nuclei in units of the proton
charge $e$. 

For neutron induced reactions, the cross section can be written as follows:
\beq
\sigma(E) = \frac{R(E)}{v},
\eeq
where $R(E)$, the $R$-factor, can be a slowly varying function of energy,
and is similar to an $S$-factor.

\subsection{Recommended Thermonuclear Rates}

\label{sect:rates}

Generally speaking, NACRE follows the standard path (outlined in the
previous section) for going from nuclear data to thermonuclear rates.
NACRE has compiled an extensive tabulation of nuclear reaction
cross section data, which is available in a very convenient
form online at the NACRE website
{\tt \url{http://pntpm.ulb.ac.be/nacre.htm}}.
For each experiment, the NACRE collaboration has organized
cross section data into useful tables, providing information on the
measured cross sections and their total error as a function of energy.
These tabulations are invaluable for the analysis of error propagation. 

It is important to have a
consistent method of analyzing cross sections and computing reaction
rates.  The data sets provided by NACRE over the relevant energy
ranges for BBN have large variations in both quality and quantity.
Thus, NACRE critically evaluates the experimental data and gives
different data sets different weights in determining rates.
The NACRE collaboration considers only charged induced
reactions,  thus leaving out several important reactions for BBN.
Therefore we have to include the neutron-induced reactions and one
deuteron-induced reaction by the SKM compilation.  The reactions 
used from each are shown in Table \ref{tab:rates}.  For
completeness we will describe the approach for the neutron-induced
reactions as well. 

Once the cross section data (and errors) are gathered, they are put
into $S$-factor or $R$-factor form.  NACRE fits these factors to
analytic functions.  For non-resonant data, a polynomial of order
2 to 3, is used.  Some data are fit from theory (e.g., NACRE
uses Kajino's \pcite{kajino} fits for
$t(\alpha,\gamma)\li7$ and $\he3(\alpha,\gamma)\be7$),
while for resonant data a Briet-Wigner or R-matrix fit is
performed (e.g., for $t(d,n)\he4$).  The fits are required to be a good
representation of the data by a $\chi^2$ analysis, taking precedence
over agreement with theory.

The evaluated cross section fits, $S_{\rm fit}(E)$ are used to determine
thermal rates (via \pref{eq:thermrate}) for a grid of temperatures, $T$,
ranging from
$T_9 = .001 - 10$, where $T = T_9\times 10^9$K.  These results are
subsequently fit to a prescribed function of $T_9$.  One should note that
the NACRE fits do not reflect the analytic approximation for the thermal
rates used by Caughlan \& Fowler (1988).  Namely, Caughlan \& Fowler put
non-resonant thermal rates in the form 
$(\sum_j c_jT_9^{j/3}) \exp(a/T_9^{1/3})$.  
NACRE modifies this approach, keeping the same
exponential dependence, but changes the prefactor from a polynomial
in $T_9^{1/3}$ to
one in $T_9$:  
$(\sum_j c^\prime_jT_9^j) \exp(a^\prime/T_9^{1/3})$.  
The main reason for the form of their fit is to
get fast convergence to the numerical data.  In some cases
(e.g. $\he3(d,n)\he4$ and $\li7(p,\alpha)\he4$) additional factors are
used to improve the fit to the numerical results.

\begin{table}[htb]
\caption{Key Nuclear Reactions for BBN}
\label{tab:rates}
\begin{tabular}{ll}
\hline\hline
Source & Reactions \\
\hline
NACRE & $d(p,\gamma)\he3$ \\
 & $d(d,n)\he3$ \\
 & $d(d,p)t$ \\
 & $t(d,n)\he4$ \\
 & $t(\alpha,\gamma)\li7$ \\
 & $\he3(\alpha,\gamma)\be7$ \\
 & $\li7(p,\alpha)\he4$ \\
SKM & $p(n,\gamma)d$ \\
 & $\he3(d,p)\he4$ \\
 & $\be7(n,p)\li7$ \\
This work & $\he3(n,p)t$ \\
PDG & $\tau_n$ \\
\hline\hline
\end{tabular}
\end{table}

As noted above, some of the rates are not provided
by NACRE.  In these cases, the SKM rates as indicated in Table
\ref{tab:rates} are used.
One of these, $\be7(n,p)\li7$,  
is a $n$-capture reaction for which a large amount of
data is available.  The deuteron-induced reaction ($\he3 (d,p)\he4$),
is fit as a charged particle reaction using the Caughlan \& Fowler
prescription, as discussed in the previous paragraph.

Several reactions deserve special mention.
As noted by SKM and emphasized recently by
Nollett \& Burles \pcite{nb},
the $p(n,\gamma)d$ reaction suffers from a
lack of data in the BBN energy range.
Also, $p(n,\gamma)d$ has only 4 data points (not available when SKM
did their study) in the relevant energy
range $\la 1$ MeV.
Fortunately, this reaction is well-described theoretically.
Here we follow both SKM and Nollett \& Burles, by adopting the
theoretical cross sections of Hale et al. (1991),
which provide an excellent fit to the four
available data points by Suzuki \pcite{suz95} and Nagai \pcite{nagai}.
Nevertheless, despite the present agreement
between theory and data, the importance of this
reaction--which controls the onset of nucleosynthesis--demands
that the theoretical cross section fit be
further tested by accurate experiment. 
We urge further investigation of this reaction.

Since SKM, Brune et al.\ \pcite{brune99} have 
added new and very precise data for $\he3(n,p)t$
(see Figure\ \ref{fig:rsfac}a).\footnote{Note
that in all figures having logarithmic vertical scales,
errors have been properly propagated to reflect the log nature
of the plot.}
This has greatly reduced the uncertainty in this reaction.
In order to use these data, we have refit the $R$
factor in the manner of SKM and Brune et al., using a third order polynomial
in $v$ and the entire world data set.  We arrive at fit parameters
very similar to those of SKM and Brune et al.,
with 
$N_A \avg{\sigma v} = \ee{7.3546}{8}(1 - 0.7757 T_9^{1/2} + 0.5376 T_9 - 0.1018 T_9^{3/2}) \ {\rm cm^{3} \ s^{-1} \ g^{-1}}$.

The weak reactions which govern $n \leftrightarrow p$ interconversion
also deserve mention. 
These reactions are a strong function of temperature, but
the rates
can be scaled to a single laboratory measurement,
the neutron lifetime $\tau_n$.
Thus, the uncertainty in the weak rates is set
by the uncertainty in $\tau_n$.
Here, we have adopted the
value recommended by the Particle Data Group,
$\tau_n = 885.7 \pm 0.8$ s. 
This value reflects the recent and very precise measurement
of Arzumanov et al.\ \cite{arz}, and has 
considerably reduced the errors of
the Particle Data Group world average from
the old determination of $886.7 \pm 1.9$ s
(Groom et al.\ \cite{PDG}).

It is also important to include the small but non-negligible corrections
to the tree-level weak rates. These 
include a number of contributions, such as radiative corrections
(Dicus et al.\ \cite{dicus};
Heckler \cite{heckler};
Esposito, Mangano, Miele, \& Pisanti \cite{emmp00a}),
nonequilibrium neutrino heating during $e^\pm$ annihilation
(Dodelson \& Turner \cite{dt}),
and finite nucleon mass and finite temperature effects
(e.g., Seckel \cite{seckel};
Kernan \cite{ker}; Lopez \& Turner \cite{lt};
Esposito, Mangano, Miele, \& Pisanti \cite{emmp00a}).
These corrections have been implemented in our code
as has been described in detail in 
Olive, Steigman, \& Walker \cite{osw00}.
As discussed there, 
the differences in the \he4 yields between our code
and that of Lopez \& Turner \cite{lt}
are $0.0001 \pm 0.0001$ for $1 \le \eta_{10} \le 10$.

\subsection{Uncertainty Limits}
\label{sect:hi/lo-def}

NACRE uses data uncertainties in two ways, for 
the evaluation of mean rates and the evaluation of errors.
For the mean (``adopted'') rates, they use the data errors to 
weight sets via a $\chi^2$ analysis.  They also use the data errors to
estimate the uncertainty range for these ``adopted'' rates.  This is
not done in a strict statistical way,  and thus is not presented as,
e.g., ``1$\sigma$'' error, but rather as ``high'' and ``low'' limits
to the thermal rates.

The ``high'' and ``low'' limits are derived in different ways for
different reactions.  In the case where there happens to be two
discrepant data sets, NACRE performs a $\chi^2$ analysis for each of
the two data sets, adopting the set with the larger $S$-factor as
their ``high'' limit, and the set with smaller $S$-factor as their
``low'' limit.  The ``adopted'' value for their mean is simply the
unweighted average of the ``high'' and ``low'' limits (e.g.,
$d(p,\gamma)\he3$).

\section{A Comparison of Nuclear Reaction Uncertainty Estimations}
\label{sect:errors}

The power of BBN theory to test and constrain cosmology and
particle physics derives from the ability to 
calculate accurately the mean values
and uncertainties in light element abundances as
a function of $\eta$.  The 
Monte Carlo analyses of BBN which provide  
these results are in turn only as good as the input 
nuclear and weak error budget.
It is thus crucial to make the most accurate estimates possible, 
in a way that is appropriate for the cross section data sets and 
the thermal rate compilation one uses.

Given the importance of the error propagation, we
will explore different methods of doing this for
the NACRE compilation.  We first examine
NACRE's own, ``high/low'' limits.  As these are
not uniformly derived, we also consider two
other error budgets, which illustrate issues
in error propagation and subsequent impact on BBN.
In this section, we will describe the 
error propagation methods, and our implementation of them.
We will then examine the BBN results using these
errors in the following section (\S \ref{sect:bbn_impact}).

\subsection{Tailoring Rates and Uncertainties for BBN}

Since the error analysis of the rates by NACRE is not done in a
strict statistical way, it becomes important to develop a general,
statistically sound method for determining an accurate mean and error
for the application to BBN.  
To do this, we will compare the NACRE theoretical $S$-factor fits
to the data sets they use.
For the theoretical fits $S_{\rm fit}(E)$, we for the most part use
the analytical forms published in the NACRE paper. 
When these were unpublished, NACRE kindly made available their fits (or in the case of 
$t(d,n)\he4$, a tabular form).\footnote{We are particularly indebted to
C. Angulo and P. Descouvrement (2000, private communication) for their help.}
These fits are valid for different energy ranges depending on
the reaction, but in all cases cover an energy band which includes
and extends beyond the energies needed for BBN.

Our goal is to assess the uncertainties
in the cross sections needed for BBN.
To do this, we will use a 
$\chi^2$ analysis to describe the goodness of fit
of the theory to the data, given the experimental errors.
However, it turns out that the 
NACRE cross section fits do not give the minimum
$\chi^2$ over the energy range for which the fits are valid.
As we will see the shape of the fits
represent the data well, but the
normalizations are not always precisely those
which minimize $\chi^2$.
This reflects the fact that 
the NACRE $S$-factors are in some cases designed to fit 
the cross section data over larger energy ranges,
and in some cases were chosen to 
be a compromise between conflicting sets of data. 
To hone the NACRE rates for use in BBN, we allow for a shift in the
overall normalization of the $S$-factor fits.  This
allows for a simple means of refining the rates for
BBN, while maintaining the shape of the cross section fits and thus
the shape of the NACRE thermal rates.
Since the fits are valid for energies beyond the BBN range,
their shapes are more strongly constrained than they would 
be if we considered only the BBN energies.

Our procedure is as follows. 
For a given reaction, we have
a set of measured $S$-factors $\{ S_{\rm obs}(E_i) \}$ measured
at energies $\{ E_i \}$, where $i \in 1,...,N$. 
The $S$-factors have errors $\{ \sigma_i \}$, while
the energies are measured with negligible error.
The measured $S$-factors are described by theoretical
fits $S_{\rm fit} (E)$, which combine theory and phenomenology
as discussed above.

\begin{table}[htb]
\caption{Reduced $\chi^2$ for Key BBN Reactions}
\label{tab:chi2}
\begin{tabular}{lrrrr}
\hline\hline
Reaction & $\hat{\alpha}-1$ & \# Points & $\chi^2_{\rm min}$ & $\chi^2_\nu$ \\
\hline
$d(p,\gamma)\he3$ & $-0.076$ & 72 & 987 & 13.9 \\
$t(d,n)\he4$ & $0.000$ & 203 & 295 & 1.46 \\
$d(d,n)\he3$ & +0.018 & 52 & 312 & 6.11 \\
$d(d,p)t$ & +0.006 & 92 & 425 & 4.67 \\
$t(\alpha,\gamma)\li7$ & +0.058 & 30 & 8.42 & 0.290 \\
$\he3(\alpha,\gamma)\be7$ & $-0.067$ & 131 & 242 & 1.86 \\
$\li7(p,\alpha)\he4$ & $-0.045$ & 126 & 374 & 2.99 \\
$\he3(n,p)t$ & 0.000 & 165 & 56.9 & 0.347 \\
$\he3(d,p)\he4$ & +0.0694  & 73 & 171 & 2.34 \\
$p(n,\gamma)d$ & $-0.012$ & 4 & 0.901 & 0.300 \\
$\be7(n,p)\li7$ & +0.010 & 65 & 38.9 & 0.608 \\
\hline
\end{tabular}
\end{table}

We allow for a different
normalization $\alpha$ between the ``true'' theory curve
and the fit curve.
That is, we write the theory curve as $\Sth(E) = \alpha S_{\rm fit} (E)$,
where we expect $\alpha$ to be near, but not necessarily equal to, unity.
Taking the data errors to be Gaussian, the data and theory
agreement is quantified by the value of chi-squared:
\beq
\label{eq:chi2}
\chi^2 = \chi^2(\alpha )
  =  \sum_i \frac{\left[ S_{\rm obs}(E_i) -\Sth(E_i) \right]^2}
                 {\sigma_i^2}
  =  \sum_i \frac{\left[ S_{\rm obs}(E_i) - \alpha {S}_{\rm fit}(E_i) \right]^2}
                 {\sigma_i^2}
\eeq
The renormalization is determined by minimizing
$\chi^2$ with respect to $\alpha$, giving the best-fit value
of $\alpha$ to be
\beq
\hat{\alpha} = \frac{\sum_i S_{\rm obs}(E_i)S_{\rm fit}(E_i)/\sigma_i^2}
                        {\sum_i [ S_{\rm fit}(E_i) ]^2/\sigma_i^2}
\eeq
which will be close to unity if the experimental data
scatters evenly around the fit curve, as expected.

The renormalizations 
appear in Table \ref{tab:chi2}.
We see that in general, the shifts are small,
less than 10\% for all cases.
The effect of renormalization can be seen graphically by
comparing the solid and long-dashed curves in Figure \ref{fig:rsfac}.
As the figure illustrates, the changes are slight--by eye,
the goodness of fit of the original NACRE fits
is similar to that of the renormalized fits. 
However, the improvement of the fits is significant enough to justify
the addition of a parameter ($\alpha$).
We will see in the next section (\S \ref{sect:central})
that the
renormalizations have a mild effect on central values
of the BBN predictions, but the shifts are nonetheless significant
since they are comparable to the level of the uncertainties.
Thus, we will show the effect on BBN of adopting the renormalizations.

\subsection{NACRE High/Low}
\label{sect:hi/lo}

As discussed above (\S \ref{sect:hi/lo-def}), for each 
reaction, NACRE presents an ``adopted''
thermonuclear rate $\lambda$, as well as a ``high''
rate $\lambda_{\rm h}$ and ``low'' rate
$\lambda_{\rm \ell}$.  Each of these is of course
a (strong) function of $T$.
NACRE describe the high/low rates as 
``lower and upper limits'' to the adopted rates, but
do not present them as statistically
defined limits, such as $1\sigma$ or $2\sigma$ ranges.
Thus, if we wish to use these limits, we must
first determine what statistical weight to
give them, and then set a prescription for
using these limits in the Monte Carlo procedure.

\begin{figure}[htb]
\begin{center}
\mbox{ \epsfig{file=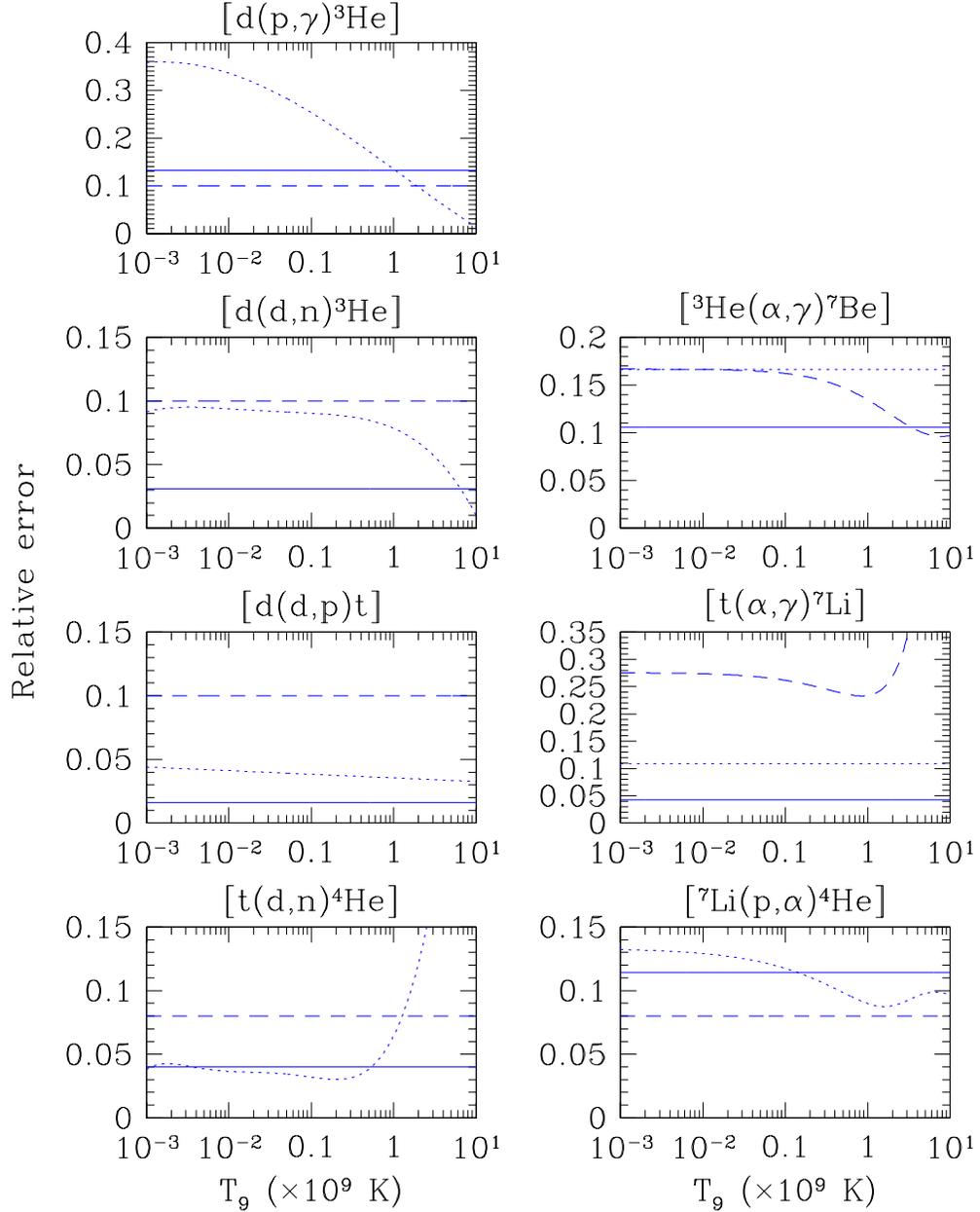,width=15cm}}
\end{center}
\vskip -1in
\caption[.]{\label{fig:high_low}
NACRE ``high/low'' fractional errors for
the seven NACRE reactions important for BBN.
Dotted curves:  high/low errors;
solid curves:  sample variance errors;
dashed curves:  SKM errors which are
shown for comparison.
}
\end{figure}

We thus turn to the question of assigning 
statistical meaning of the NACRE high/low limits.
The uncertainties are conveniently quantified 
by the 
{\em fractional} or relative error $\delta_i = |\lambda_i/\lambda - 1|$,
with $i=$ high and low
In principle, for a given reaction rate
the high and low errors need not be equal,
but in practice we find that the high and low 
fractional errors are nearly identical
(certainly within our desired accuracy)
over the entire temperature range.  Thus, we can write
the high/low limits as 
$\lambda_i(T) = \lambda(T) \ [1 \pm \delta(T)]$,
where $\delta = \delta_\pm$.  This simplifies
the analysis slightly.  
Figure \ref{fig:high_low} plots the fractional errors
$\delta(T)$ for the seven
NACRE reactions important for BBN.
For comparison, we have also plotted fractional errors given
in the SKM compilation.\footnote{In the BBN code,
for the three ``SKM-only'' strong
reactions (Table \ref{tab:rates}) as well as for $\he3(n,p)t$, 
we apply the same analyses as in the case of the NACRE rates,
with the exception that NACRE's high/low error
estimate is replaced
by the SKM recommendation.}  
Several trends are apparent in the NACRE curves.  
We see that
in general, the NACRE high/low fractional errors
vary with temperature; this is in contrast with
the SKM rates, for which only 2 of the 12 rates are
assigned a temperature-dependent fractional error.

Comparing the high/low limits with the $1\sigma$ SKM errors, 
we see that the
errors are of the same order of magnitude.
There is no clear trend as to which compilation has
the larger errors.  For example, the NACRE errors for 
$t(\alpha,\gamma)\li7$ are
lower than the SKM rates (due in part to the
addition of new, accurate data by
Brune et al.\ \cite{brune}), while
for the
mirror reaction 
$\he3(\alpha,\gamma)\be7$, the NACRE limits are
larger than those of SKM.
For the purpose of comparison with our more rigorous results which 
follow,
we have taken the simple
approach of assigning the NACRE high/low limits
to be $1\sigma$ errors.
Given the variations in the NACRE rates and their comparison with the
SKM variations, this approach is certainly qualitatively
correct.
A similar approach was taken
by Vangioni-Flam, Coc, \& Cass\'{e} \pcite{vcc},
who examined the impact of the NACRE rates and ran cases
using the high/low limits to compare with the central rates.

While the NACRE assignment of limits to their rates
invites quantification in a full Monte Carlo code,
the somewhat arbitrary nature of the
specific assignment of $1\sigma$ errors to 
the limits leaves question as to the robustness of
the results. While here we have assumed a Gaussian distribution for
these assignments, other choices such as a log-normal distribution
are also possible (Vangioni-Flam, Coc, \& Cass\'{e} \pcite{vcc2}).
Thus, we now turn to other error assignments which are derived from
the underlying cross section data.

\subsection{Minimal Uncertainties:  $\Delta \chi^2 = 1$}
\label{sect:delta-chi2}

To derive errors for  NACRE rates in a more
statistically well-defined way, we now
perform our own analysis of data
and propagation of errors.
In this section, we will
treat all cross sections errors, from all data sets, as
independent.  We thus ignore any correlations between
the measurements, even within a given experimental run.
Clearly, this is not realistic, but this procedure has
the virtue of simplicity and  will give the limiting case
of the smallest possible error achievable with
nuclear data.  This will also serve as a useful
starting point for a more realistic analysis in
the next subsection.

We will compute the variance in the $S$-factor in the
form of a fractional error.
This is also an approximation, but simplifies the analysis
and the limit that results.
Physically, this corresponds to the assumption
that the large amount of data 
determines the {\em shape} of the curves quite
well, so that the dominant uncertainty is
not in the shape but in the overall {\em normalization}.
The data sets used are the same as NACRE; the
online versions include data and
errors which is greatly helpful in this process.

We can determine the fractional {\em error} in the theory
fit by varying the normalization until
$\chi^2 = \chi^2_{\rm min} + 1$,
which occurs when  
\beq
\alpha_1 = \hat{\alpha}
  \pm \frac{1}{\sqrt{\sum_i S_{\rm fit}(E_i)^2/\sigma_i^2 }}
\eeq
Thus, we have a $1\sigma$ variation when $\alpha$
is changed from its minimizing (renormalized) value
by a fractional error 
\beq
f = \alpha_1/\hat{\alpha} - 1
\eeq
The theory-data fit thus has a $\pm 1\sigma$ error
$\sigma_{\rm th} = f \Sth = \hat{\alpha} f S_{\rm fit}$.

We have found these ``$\Delta \chi^2 = 1$'' errors 
for the 11 strong reactions; these
appear in Table \ref{tab:errors}.
In the Table, we see that the errors are very small, $\la 1\%$.
This can be seen to follows from the large number of data points.
Of course, we have in this case we have assumed
each measurement in each experiment to be independent
of all others.  We have thus ignored the correlations
among the errors.  These problems are highlighted
when we examine the questions of goodness of fit.

\begin{figure}[htb]
\begin{center}
\mbox{ \epsfig{file=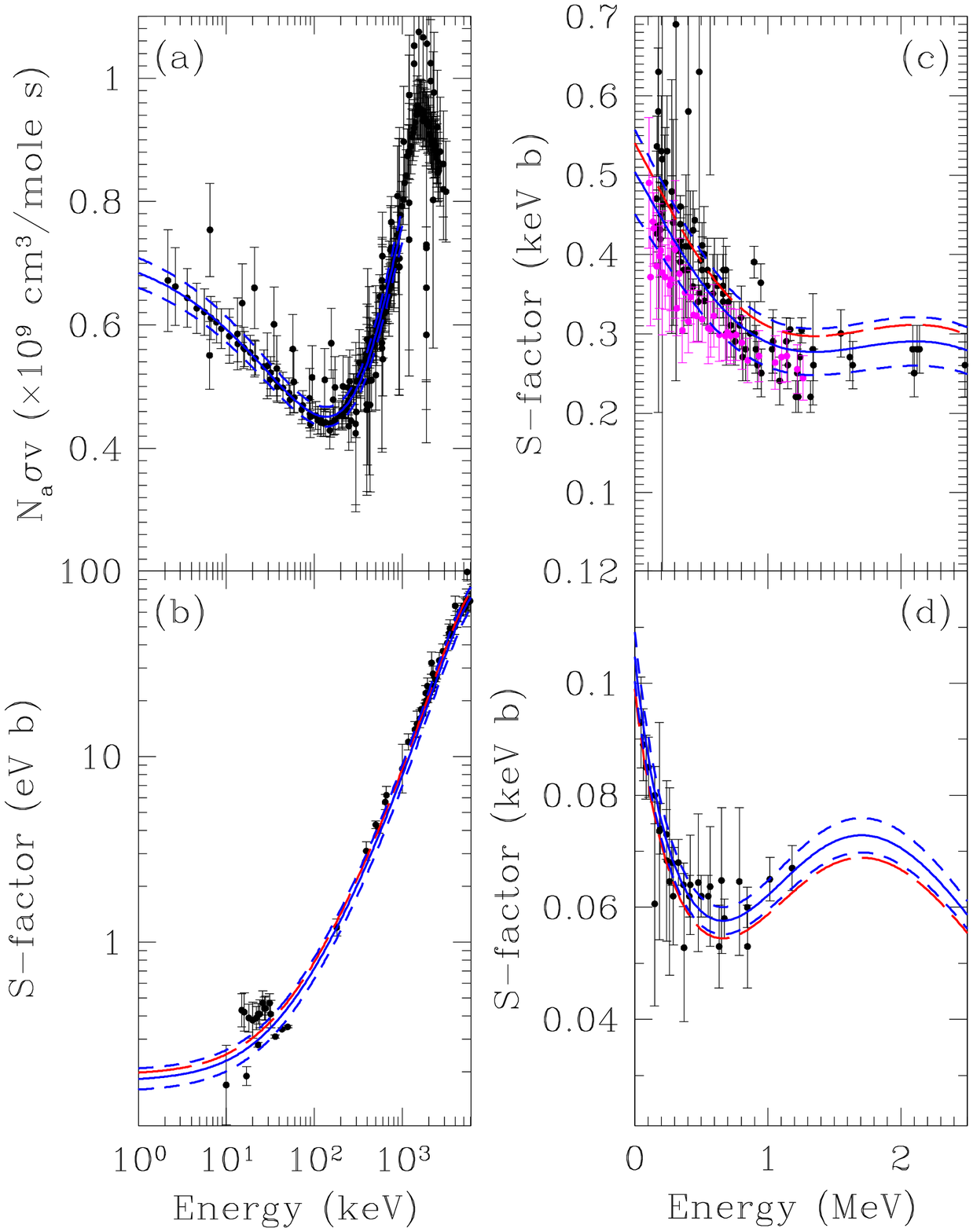,width=15cm}}
\end{center}
\vskip -1in
\caption[.]{\label{fig:rsfac}
Experimental data and $R$- and $S$-factor fits
for four key reactions.
The {\em solid curves} are our renormalized fits, with our sample
variance error bands
given by the short dashed curves.
For the $S$-factor curves, data are given in NACRE,
whose raw fits are given by the {\em long dashed curves}.
a) The $R$-factor for $\he3(n,p)t$.  The data
are those of SKM with the addition of Brune et al.\ \cite{brune99}.
The fit is ours, as described in \S \ref{sect:rates}.
b)
The $S$-factor data for $d(p,\gamma)\he3$.
c)
The $S$-factor data for $\he3(\alpha,\gamma)\be7$.
d)
The $S$-factor data for $t(\alpha,\gamma)\li7$.
}
\end{figure}

The goodness of fit is quantified by $\chi^2_\nu = \chi^2/\nu$,
the $\chi^2$ per degree of freedom with $\nu = N-1$. 
Table \ref{tab:chi2} gives the
reduced $\chi^2$ for the 11 key strong reactions.
{}For normally distributed and
correctly estimated
errors, one expects $\chi^2_\nu$ to lie
close to 1.
In fact, this is not what is found for any of the reactions.
In four cases,
the reduced $\chi^2$ is significantly less than 1,
while 
in seven cases $\chi^2$ significantly exceeds 1.
In Figure \ref{fig:rsfac}d, we see the data and fit for
$t(\alpha,\gamma)\li7$, an example of a reaction with 
$\chi^2_\nu <1$.  In this case, we see that all data are
within $1\sigma$ of the fit, and indeed some lie essentially 
on the fit, leading to the small reduced $\chi^2$ and suggesting
that the experimental errors may be overly conservative in this case.

For the cases in which $\chi^2_\nu > 1$, 
we follow the practice of the Particle Data Group and we increase
the fractional error by an amount $\sqrt{\chi^2_\nu}$. 
This scaling encodes the presumption that at least one
of the experiments has underestimated its uncertainties.  
Where the data are discrepant, this
is most assuredly the case.
For the cases in which the $\chi^2_\nu < 1$, no correction is made.
 
Let us examine more closely the large $\chi^2_\nu$ cases,
for which the goodness of fit is poor.
A particularly egregious case is $d(p,\gamma)\he3$,
with $\chi^2_\nu = 13.9$, which appears in
Figure \ref{fig:rsfac}b.
The reason for the poor fit is apparent:
the data for this reaction are internally
inconsistent in the 0.01 to 0.05 MeV range.  Specifically, the recent
measurements of Schmid et al.\ \pcite{schmid}
do follow the fit trend, but the older and less
precise data of Griffiths, Lal, \& Scarfe \pcite{gls}
lie above these data, outside of the range of the quoted
errors.  In our fitting procedure, the small errors in the
Schmid et al.\ \pcite{schmid} data dominates the fit normalization,
but the systematic offset of the Griffiths, Lal, \& Scarfe \pcite{gls}
data leads to a poor $\chi^2$.

Another case with a high $\chi^2_\nu$ is
$\he3(\alpha,\gamma)\be7$, which appears 
in Figure \ref{fig:rsfac}c.  
This reaction has been well-studied in the context of
solar models, where it controls the
intensity of \be7\ neutrinos
and thus is of great importance for
the solar neutrino problem.
Studies of this reaction are 
reviewed by Adelberger et al.\ \pcite{adel},
who note the discrepancy between $\gamma$-capture and \be7 activity
measurements.  Although the discrepancies between the data sets are
not as glaring as in the $d(p,\gamma)\he3$ case, 
the systematic offset between data sets is nevertheless clear.

The source of the systematic error traces
back to the difficulty in making an absolute cross section
measurement.  While a given experiment can accurately determine 
the relative values between different energies, 
measuring the absolute cross section values involves
calibration of beam intensities, beam energy losses in the
target, detection efficiencies, solid angle uncertainties, etc. 
Thus, the absolute cross sections from each experiment
carry a normalization error, which is common to all points
from the experiment.  As emphasized by 
Burles, Nollett, Truran, \& Turner \pcite{bntt}
and Nollett \& Burles \pcite{nb}, these normalization errors 
in the cross sections affect all of the BBN reactions,
and are properly treated as correlated errors.

Thus, our assumption of independent errors,
and the $\Delta \chi^2 = 1$ method, while simple
and illustrative, has greatly {\em underestimated}
the true error budget, in which correlations play
an important and even dominant role.
We now turn to a method which 
includes these effects
and treats the data in a more uniform manner.

\subsection{Sample Variance}

We now seek a new error estimator which accounts for correlated
errors due to absolute normalization
uncertainties in the cross section measurements.
These dominate the uncertainties for many reactions.
Furthermore, these uncertainties are not reduced by taking
more data within a given experiment; rather, each 
experiment effectively represents {\em one} attempt to measure
the absolute normalization, independent of the number of
points measured in the experiment.  Thus, the normalization errors
for a given reaction are only 
lowered by combining many independent experiments.
Consequently, the overall error budget in the cross sections, and thus
in the inferred thermonuclear rates, is much larger.

We note that the systematic errors in absolute cross section
normalization are quoted as percentages.
Thus, the use of fractional errors is justified when
these errors dominate.  We will thus cast
our uncertainty determinations in the form of a
fractional or normalization uncertainty (as in the previous
section), but now with more justification.

In constructing an error estimate which accounts for
correlated normalization errors, 
the data sets with discrepant normalizations will
provide a useful test case.  Our goal is to
find a well-defined procedure which will give
errors that automatically account for
systematic offsets among data sets, 
leading to an error range which
accommodates both sets.
We therefore 
adopt a (fractional) error determined
by the weighted sample variance, as follows.
The notation is the same as that introduced
in the previous section.  In particular,
we again adopt a theory
normalization $\alpha$ given by
$\Sth = \alpha S_{\rm fit}$,
and use the data to determine the value
of $\alpha$ and its error.
The mean value $\hat{\alpha}$ of the normalization is determined again
by $\chi^2$ minimization.
Given this, we can find the fractional 
difference 
\beq
\delta_i = \frac{S_{\rm obs}(E_i)-\Sth(E_i)}{\Sth(E_i)}
  = \frac{S_{\rm obs}(E_i)}{\hat{\alpha} S_{\rm fit}(E_i)} - 1
\eeq
between each data point and the theory,
and the fractional error 
\beq
\epsilon_i = \sigma_i/\Sth(E_i) = \sigma_i/\hat{\alpha} S_{\rm fit}(E_i)
\eeq
in each data point.
Given these quantities, the
theory normalization $\hat{\alpha}$ which minimizes $\chi^2$ 
(eq.\ \ref{eq:chi2})
is also the value of $\alpha$ which  guarantees that the weighted average
theory-data difference satisfies
$\avg{\delta} = \sum \delta_i/\epsilon_i^2 /\sum 1/\epsilon_i^2 = 0$.
The error, as measured by the 
weighted sample variance $\sigma_\alpha$, is now determined as
the weighted average of $\delta_i^2$:
\beq
\label{eq:sampvar}
\sigma^2_\alpha  
  = \frac{\sum_i \delta_i^2/\epsilon_i^2}
         {\sum_i 1/\epsilon_i^2}
  = \frac{\sum_i \left[ S_{\rm obs}(E_i)-\Sth(E_i) \right]^2/\sigma_i^2 }
         {\sum_i \Sth(E_i)^2/\sigma_i^2}
\eeq
Note that if a typical data point $S_{\rm obs}$ lies
$\sigma$ away from the theory point $\Sth$, then 
$\sigma_\alpha \sim \sigma/\Sth$, the fractional error.
However, since the cross sections are a function of energy and are not
constant, the weighting will favor points which have
a small fractional error $\sigma_i/\Sth(E_i)$.

We now turn to the behavior of 
eq.\ \pref{eq:sampvar} in the case
of discrepant data sets.
For a simple but illustrative example, consider a reaction
for which there are two data sets which are inconsistent with
each other.  Let each data set contain $N$ points, with comparable errors.
Let the first set have measurements with 
$S_{\rm obs} (E_i) = \alpha_1 S_{\rm fit}(E_i) + \delta_i$
for $i \in (1,\ldots,N)$,
so that the data scatters about the fit curve
with a normalization discrepancy (systematic error) $\alpha_1$,
and the errors which scatter as $\avg{\delta_i} =0$ and
$\avg{\delta_i^2}=\sigma_i^2$.  
Similarly, the second data set has points with
$S_{\rm obs} (E_i) = \alpha_2 S_{\rm fit}(E_i) + \delta_i$
for $i \in (N+1,\ldots,2N)$,,
where $\alpha_2 \ne \alpha_1$ indicates a systematic discrepancy,
and  $\avg{\delta_i^2}=\sigma_i^2$.  
Finally, as described above, 
we allow for a theory normalization error:
$\Sth = \alpha S_{\rm fit}$. Then the expected $\avg{\chi^2}$ is 
minimized for the normalization
\beq
\label{eq:renorm}
\hat{\alpha} 
  = \frac{\sum_{i=1}^{N} S_{\rm fit}{(E_i)^2}/\sigma_i^2}
         {\sum_{i=1}^{2N} S_{\rm fit}{(E_i)^2}/\sigma_i^2} \; \alpha_1
  + \frac{\sum_{i=N+1}^{2N} S_{\rm fit}{(E_i)^2}/\sigma_i^2}
         {\sum_{i=1}^{2N} S_{\rm fit}{(E_i)^2}/\sigma_i^2} \; \alpha_2
\eeq
i.e., a weighted average of the two factors $\alpha_j$.
This result is of course quite reasonable.
In this case, we have
\beq
\avg{\chi^2(\hat{\alpha}) }
   = 2N 
   + (\Delta \alpha/2)^2\sum_{i=1}^{2N} \frac{S_{\rm fit}{(E_i)^2}}{\sigma_i^2}
\eeq
We see that $\chi^2$ differs from the number $\nu = 2N-1 \sim 2N$
of degrees of freedom by an amount which depends on
$\Delta \alpha$, the difference in the systematic error.
In this case, therefore, the degree to which $\chi^2$ exceeds $\nu$,
or $\chi^2_\nu$ exceeds 1, is just a measure of the 
systematic discrepancy between the data sets.  Again, this is as expected.

For the special case when $S_{\rm fit} = S_0$ and
$\sigma_i = \sigma_0$, both constants independent of energy,
eq.\ \pref{eq:renorm}
reduces to just $\hat{\alpha} = (\alpha_1 + \alpha_2)/2$, the
simple average.
Furthermore, in this case the weighted sample variance of
eq.\ \pref{eq:sampvar} takes the simple form
\beq
\sigma_\alpha^2 = (\sigma_0/S_0)^2 + (\Delta \alpha/2)^2
\eeq
where $\Delta \alpha = \alpha_2-\alpha_1$.
Again, this result is quite reasonable:  the fractional error in the
normalization is the quadrature sum of
a typical fractional  measurement error $\sigma_0/S_0$
and the systematic error as estimated by $(\alpha_2 - \alpha_1)/2$, the
difference in normalizations.  Thus, when
the systematic error dominates, then $\sigma_0/S_0 \ll \Delta \alpha$ 
we have an overall fractional error
$\sigma_\alpha \simeq \Delta \alpha/2$, so
that the range $\alpha = \hat{\alpha} \pm \sigma_\alpha$
corresponds precisely to the range $\alpha_1 \le \alpha \le \alpha_2$.
This is precisely what one expects if the systematic error dominates;
our expression automatically gives an answer which spans the range
between the discordant data sets.
On the other hand, if the systematic error is negligible, then
we have $\sigma_0/S_0 \gg \Delta \alpha$, and
$\sigma_\alpha \simeq \sigma_0/S_0$, a typical measurement error.
In this limit, then, the fractional error is such that it encloses
the $1\sigma$ error band about the (normalized) fit curve.

\begin{table}[htb]
\caption{Reaction Errors}
\label{tab:errors}
\begin{tabular}{lcc|c}
\hline\hline
  & Minimal Error& Sample Variance & SKM fractional \\
Reaction & $f$ & $\sigma_\alpha$ & error \\
\hline
$d(p,\gamma)\he3$ &0.00420 &0.132 &0.10 \\
$t(d,n)\he4$ &0.00233  &0.0401 &0.08\\
$d(d,n)\he3$ &0.00176  &0.0310 &0.10\\
$d(d,p)t$ &0.00077  &0.0159 &0.10\\
$t(\alpha,\gamma)\li7$ &0.0145  &0.0421 & $\sim 0.23-0.30$ ($T$-dep)\\
$\he3(\alpha,\gamma)\be7$ &0.0068 &0.106 & $\sim 0.10-0.17$ ($T$-dep)\\
$\li7(p,\alpha)\he4$ &0.0059 &0.114 &0.08\\
$\he3(n,p)t$ &0.00467 &0.03523 &0.10\\
$\he3(d,p)\he4$ &0.00699 &0.0915 &0.08\\
$p(n,\gamma)d$ &0.00324 &0.0445 &0.07\\
$\be7(n,p)\li7$ &0.00621 &0.0387 &0.09\\
\hline
\end{tabular}
\end{table}

Table \ref{tab:errors} shows $\sigma_\alpha$ for the
11 strong BBN reactions.
We see that these results are either comparable
to or smaller than the SKM results.
The largest uncertainties appear when incompatible data sets
are present.  As expected, the systematic offsets
are accounted for by an increase in the errors.

\subsection{Other BBN Error Studies}

Other studies have been made
which determine errors in the nuclear inputs for
BBN and propagate them to derive
uncertainties in BBN abundance predictions.
The SKM compilation of nuclear uncertainties
has been very widely used for BBN Monte Carlo 
studies.
This work was extremely influential,
as it was the first to 
catalog the most important BBN reactions,
 to systematically collect, 
and display the available cross section data for them,
and to derive a set of errors for them.
The explicit table of uncertainties (given as
fractional errors)
made the SKM work very useful, as these
that could be readily input into BBN codes.

An important recent study takes a very different approach
to the BBN error budget.
Burles, Nollett, Truran, \& Turner \pcite{bntt} and
Nollett \& Burles \pcite{nb}
made an extensive 
Monte Carlo calculation of the light element abundances
and their uncertainties.
In this work, the $R$- and $S$-factors for the 11 key strong
reactions are generated
from fits to simulated data sets.
These mock data are
drawn randomly, on the basis of the actual
data.
For each realization of the nuclear data,
the thermonuclear rates are computed
via eq.\ \pref{eq:thermrate}, and run in the BBN code.
Care is taken to allow for correlations due
to normalization errors. We note however, that there is a 
danger in this procedure, as the resulting adopted $S$ and $R$-
factors may not have a dependence as a function of energy 
as expected from theory.  Indeed, some cases, such as
$^7$Li(p,$\alpha$)$^4$He, show a behavior which is probably not
physical and is simply a result of the available data at particular
energies\footnote{This was pointed out by A. Coc in his talk
at Cosmic Evolution, Paris, Nov.\ 2000 (Coc, Vangioni-Flam \& Casse 2001).}

Thus, Nollett \& Burles do not directly report a mean and error for
the thermonuclear rates (though they do produce mean values and
errors for the $R$- and $S$-factors). 
Instead, they report the final results for the abundance mean
values and errors, as a function of $\eta$.
Finally, we note that 
Nollett \& Burles adopt a 
neutron lifetime $\tau_n = 885.4 \pm 2$ s,
which is slightly lower than the current world
average of the Particle Data Group (Groom \etal 2000),
$885.7 \pm 0.8$ s. 
This difference
affects mostly \he4, with a small drop
$\Delta Y_p \simeq -0.0001$ for the Burles \& Nollett value,
although the error used here is considerably smaller.

\begin{figure}[htb]
\begin{center}
\mbox{ \epsfig{file=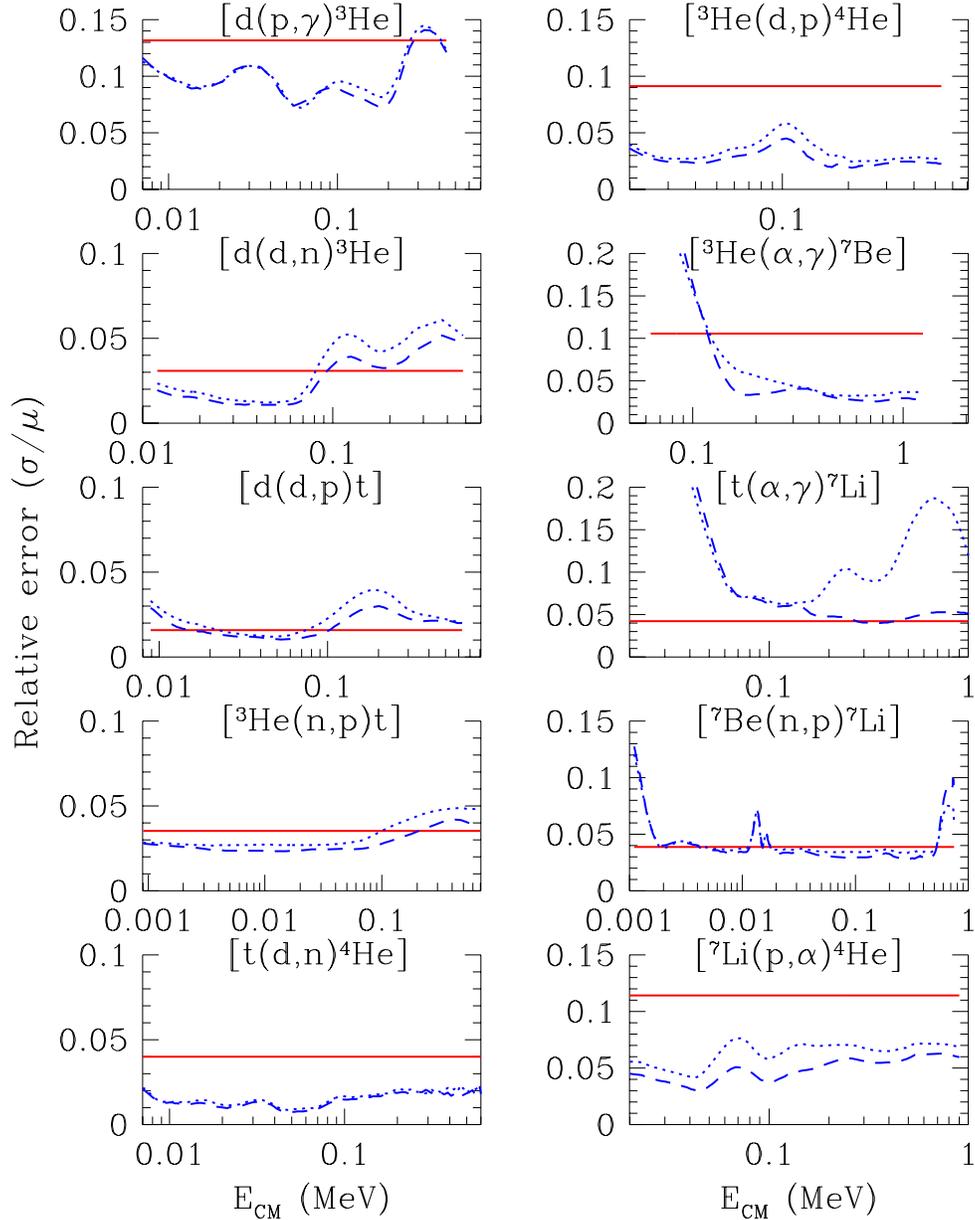,width=15cm}}
\end{center}
\vskip -1in
\caption[.]{\label{fig:err-compare}
Fractional $1\sigma$ errors in the $R$- and $S$-factors
for sample variance analysis, as
well as Nollett \& Burles \pcite{nb}.
Solid curves:  sample variance limits; 
dashed curves: Nollett \& Burles upper limits;
dotted curves: Nollett \& Burles lower limits.
The $p(n,\gamma)d$ reaction is not shown,
as Nollett \& Burles adopt a constant $1\sigma$ fractional error
of 5\% in this case.
}
\end{figure}

To compare our nuclear error budgets with those of SKM
and Nollett \& Burles  \cite{nb}, we have plotted
fractional errors in the $R$- and $S$-factors 
for 10 of the 11 strong reactions.\footnote{For legibility,
we have omitted $p(n,\gamma)d$, for which
Nollett \& Burles adopt a constant fractional
error of 10\% at the 95\% CL. The corresponding SKM uncertainty is 14\%
and our sample variance error is $\sim$ 9\% (see Table 2).}   These appear
in Figure \ref{fig:err-compare}.
We see that our sample variance uncertainties
are always tighter than those of SKM.
On the other hand, our results are almost always
larger than or about equal to the errors found by
Nollett \& Burles.\footnote{Note that the 
plotted Nollett \& Burles errors
are 1/2 of their 95\% CL uncertainties.}  
It is interesting that, with the exception
of the two $(\alpha,\gamma)$ reactions,
the Nollett \& Burles fractional errors are roughly
constant with energy.  
In four of these cases, our values
agree well with the average of the range spanned by Nollett \& Burles.
For $d(p,\gamma)\he3$, our value is slightly higher,
due to method's slightly more conservative treatment of the
discrepant data for this reaction as is the case for $\li7(p,\alpha)\he4$.
In the mirror reactions of $t(d,n)\he4$ and $\he3(d,p)\he4$,
the reactions proceed through a resonance, and thus
the errors are very sensitive to the goodness of fit
for the NACRE and SKM curves, which is not an issue for
the empirical approach of Nollett \& Burles.

\section{The Impact on BBN}
\label{sect:bbn_impact}

We have now adopted a set of thermonuclear
rates using the NACRE compilation (\S \ref{sect:nacre}),
and we have presented several estimates
of the uncertainties in theses rates (\S \ref{sect:errors}).
In this section, we will put these rates into the BBN
code, and compute the light element
abundances and their errors
as a function of $\eta$.  
We can then quantify
the impact of the NACRE
compilation on the
mean values of the abundances, and on their uncertainties.

The implementation of the Monte Carlo is standard
and well-described in, e.g., 
Krauss \& Romanelli \pcite{kr} and
SKM.
Briefly, a grid of $\eta$ 
is chosen, with values
in the $(1-10) \times 10^{-10}$ range. 
At each $\eta$, 
the code is run 1000 times.
For each run, we generate 12 Gaussian random numbers
$z_i$ with zero mean and unit variance; these
are used to choose the thermonuclear reaction
rates $\lambda_i$ 
in terms of their mean values $\lambda_{i,0}$ and
fractional errors  $f_i = \delta \lambda_i/\lambda_{i,0}$:
\beq
\lambda_i(T) = \lambda_{i,0}(T) \ \left[ 1+ z_i f_i(T) \right]
\eeq
Where $f_i(T)$ is constant for the minimal uncertainty
and sample variance cases, and $T$-dependent for the high/low case.

At each $\eta$, 
the central value of the light element
abundance $\hat{y_i}$ is taken to be the mean of the
1000 values from the runs, and
the standard deviation $\sigma(y_i)$ is taken from
the sample variance of the 1000 runs.
A technical note:  
the sample variance itself has an expected deviation 
of $\delta \sigma^2 \sim \sigma^2/N$ from 
the true error, so that our uncertainty estimates have 
a statistical error of $1/\sqrt{N} \simeq 3\%$.
To allow for this, we increase the 
computed sample variance by 3\%; in every case the effect is small.

\subsection{NACRE Central Values}
\label{sect:central}

\begin{figure}[htb]
\begin{center}
\vskip -0.5in
\mbox{ \epsfig{file=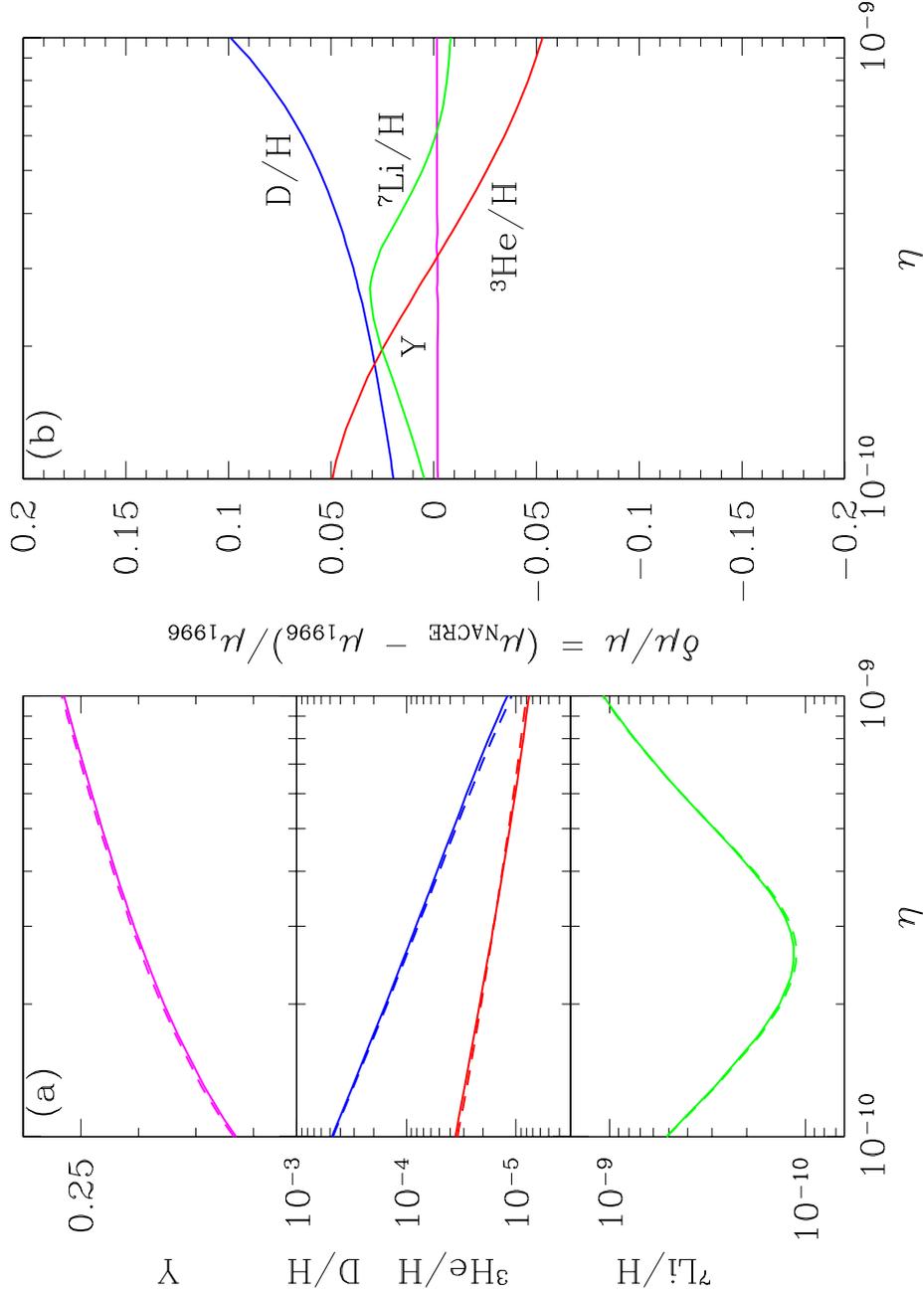,width=15cm}}
\end{center}
\vskip -0.5in
\caption[.]{\label{fig:nacre_vs_skm_noren}
Light element abundances as a function of $\eta$,
using NACRE rates without renormalization.
a)
The solid curves are the central values of the NACRE predictions,
dashed curves are those of SKM.  The two are extremely close,
so that they often appear to overlap.
b)
Percent difference $100(y_i^{\rm NACRE}/y_i^{\rm 1996} - 1)$
between the 
NACRE and Hata \etal (1996) central values.
}
\end{figure}

It is important to first compare the direct impact of the NACRE rate
compilation on BBN predictions.  To do this we will use the expanded BBN
code of Thomas
\etal (1993),
which was based on the Kernan (1993) code, with the Monte Carlo
implemented by Hata \etal (1996). The uncertainties in the 1996 Monte
Carlo were derived from SKM. The central values of the light element
abundances appear in Figure
\ref{fig:nacre_vs_skm_noren}a, which shows results for the NACRE rates
along with the Hata \etal (1996) results. (the solid and dashed
curves).
We see that the two predictions are almost identical
to each other.  
The differences between the two cases are plotted in 
Figure \ref{fig:nacre_vs_skm_noren}b as a percentage of the 1996
predictions. We see that the changes are small:  the largest shift is a
10\% rise in D/H at $\eta = 10^{-9}$, and that below $\eta = 3 \times
10^{-10}$, all shifts are smaller than 5\% in magnitude.

These results
confirm the findings of Vangioni-Flam, Coc, \& Cass\'{e} \pcite{vcc}:
the central values of the BBN predictions are insensitive
to the reaction network choice.
It is both non-trivial and comforting that different rate tabulations,
with different functional forms, yield the same central values.
This agreement
reaffirms that basic predictions of BBN are robust.

\begin{figure}[htb]
\begin{center}
\vskip -0.5in
\mbox{ \epsfig{file=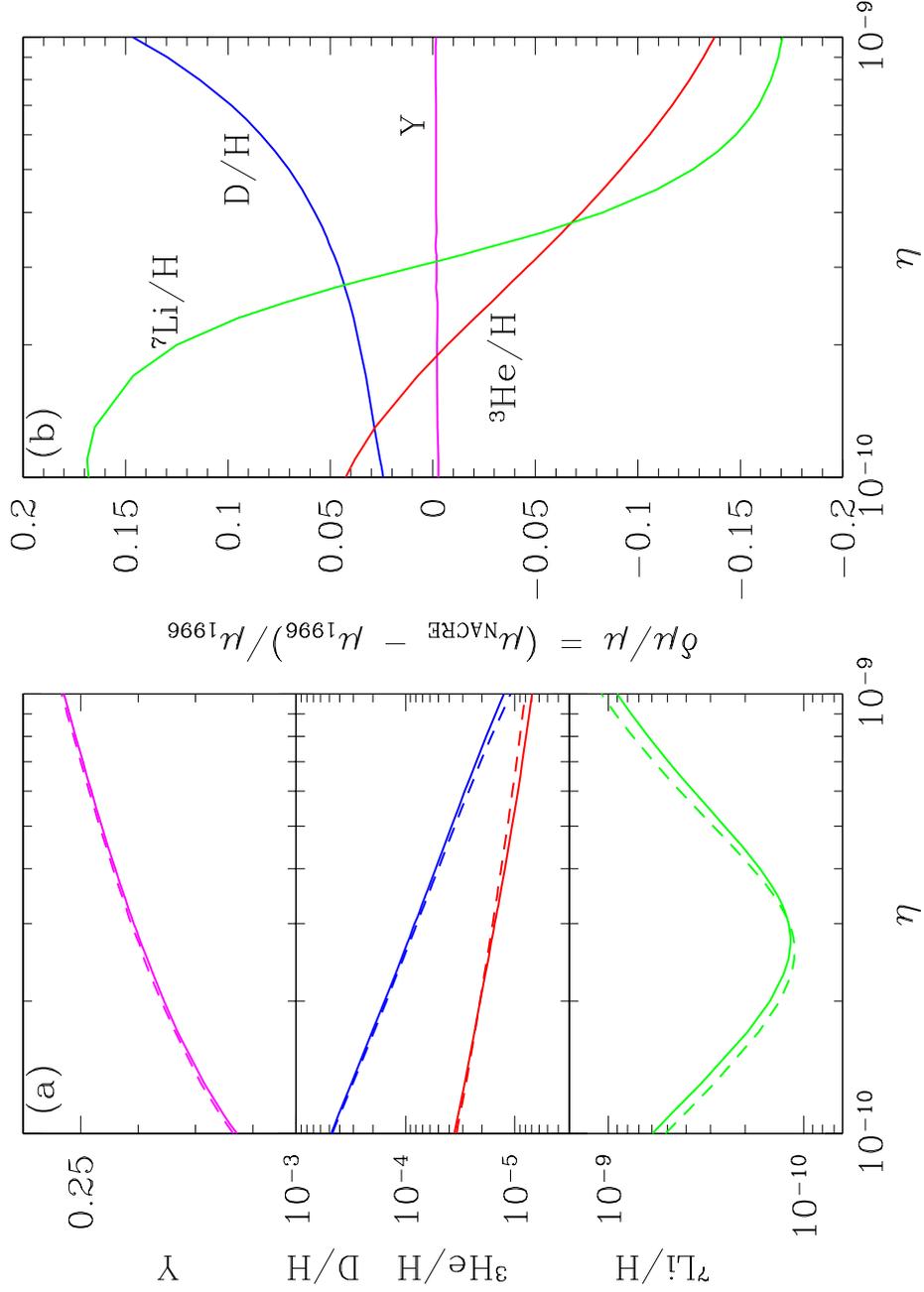,width=15cm}}
\end{center}
\vskip -0.5in
\caption[.]{\label{fig:nacre_vs_skm_ren}
As in Figure \ref{fig:nacre_vs_skm_ren}, for light element
predictions with renormalization.

}
\end{figure}

Although the 
central light element predictions
of the two rate compilations do not differ
significantly, a larger effect is found
by renormalizing the rates to reflect the data in
the BBN energy range.
Figure \ref{fig:nacre_vs_skm_ren}a
shows the predictions when the rates are renormalized
as described in the previous section.
We see that central values do change, 
at levels larger than those discussed above.
Figure \ref{fig:nacre_vs_skm_ren}b shows the percent difference of
these predictions with those of Hata \etal (1996).
We see that 
the shifts are $\ll 1\%$ for $Y_p$, 
at a level of $\le 15\%$ for D and \he3,
and vary the most for \li7, which runs between $\pm 20\%$.
The shifts trace back to the rates which have the
largest renormalizations.
In particular, the $d(p,\gamma)\he3$ rate
drops by 7.6\% with respect to NACRE
due to the preferential weighting of the
Schmid et al.\ \pcite{schmid}
data with their smaller errors (Figure \ref{fig:rsfac}d).
This change alone is predominantly
responsible for the shifts in D, \he3, and \li7 at
$\eta \ga 3 \times 10^{-10}$;
the $\he3(d,p)\he4$ +6.9\% renormalization is
responsible for most of the rest of the shift in this regime.
At smaller $\eta$, there is a rise in \li7; this is
predominantly due to the combination of the
increase in \li7 production via
a 5.8\% rise in the $t(\alpha,\gamma)\li7$ rate,
and a decrease in \li7 destruction due to
a 4.5\% lowering of the $\li7(p,\alpha)\he4$ rate.
We note that these shifts in the central values of the 
predicted abundances are well within the calculated uncertainties
discussed below.

\subsection{NACRE High/Low Errors}

\label{sect:bbn-hi/lo}

\begin{figure}[htb]
\begin{center}
\vskip -0.5in
\mbox{ \epsfig{file=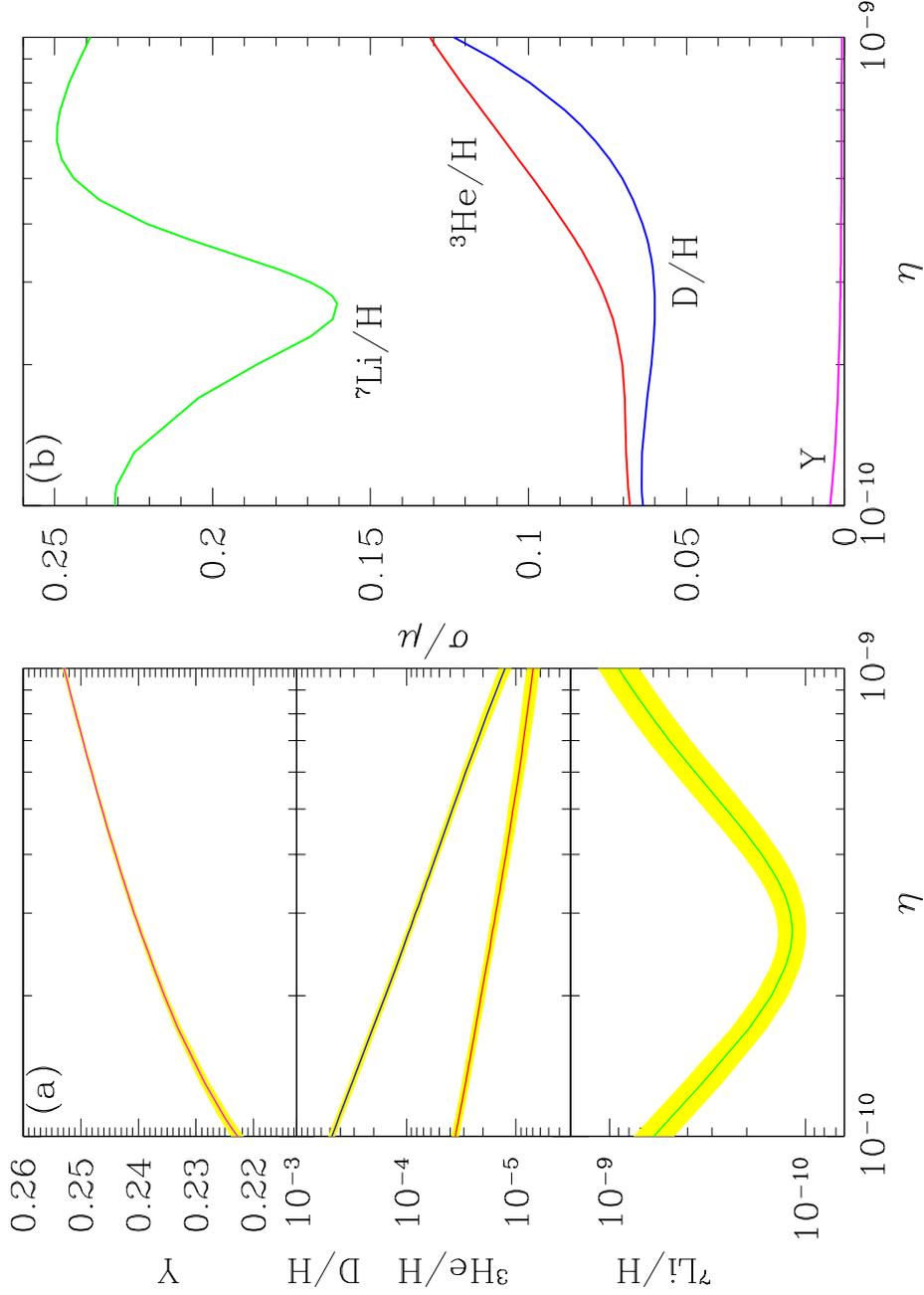,width=15cm}}
\end{center}
\vskip -0.5in
\caption[.]{\label{fig:bbn-hi/lo}
Light element abundance predictions, and their uncertainties,
as a function of $\eta$. 
a)
The solid curves are the (renormalized) NACRE abundance predictions,
and the broken curves are the $1\sigma$ Monte Carlo errors for 
the NACRE high/low error estimates.
b)
The fractional errors in the 
light element predictions,
for the NACRE high/low estimates
as in (a)
(solid curves).
The fractional errors plotted are
$\sigma_i/\mu_i$, where $\mu_i$ is the mean
value of abundance $i$, and $\sigma_i$ is
its error. 
}
\end{figure}

The impact of NACRE's high/low errors
(\S \ref{sect:hi/lo}) on the light element abundances
is seen in Figure \ref{fig:bbn-hi/lo}a.
The logarithmic nature of the standard
abundance plot, and the rapid variation of
the trace elements, makes it difficult to
assess and compare the errors.
Thus we have
also plotted the fractional errors
$\sigma_i/\mu_i$ 
in Figure \ref{fig:bbn-hi/lo}b.  
We see that overall, the NACRE high/low errors
are quite comparable to those found using the SKM estimates.
The most significant effect
is the reduction in the \li7 errors,
which is due to the lower uncertainties in
the mass-7 production reactions
$t(\alpha,\gamma)\li7$ and
$\he3(\alpha,\gamma)\be7$.
Here, the errors are a strong function of $\eta$ owing to the
\li7 and \be7 production channels, and the errors are comparable
to those of SKM at $\eta_{10}=10^{10} \eta$ near 3.
However, the error on Li is reduced by as
much as $\sim 25\%$ for $\eta_{10}$ near 1 or 10,

We confirm here the key result that the
theoretical errors in BBN are still significant.
The errors in D and \li7 in particular are only marginally
smaller than the accuracy of current or near-future observations.
Thus, improvements in the nuclear inputs to BBN can still
have an important effect on cosmology, as we will
discuss below (\S \ref{sect:eta}).

It is important to note that our
results differ somewhat from those of 
Vangioni-Flam, Coc, \& Cass\'{e} \pcite{vcc},
due to differences in approach.
Vangioni-Flam et al.\ used the high/low errors
to examine the effect on BBN of each reaction
individually.  They then found the abundances
when all the reactions
are set to their high limits, then when
all reactions are set to their low limits.
This procedure is a simple and rapid means of
gauging the net effect of the uncertainties.
However, as Vangioni-Flam et al.\ note,
this can lead to compensating effects
(e.g., since both the \li7 destruction
and production rates are raised, there is
a degree of cancellation in the effect of the
errors on the final
abundances).
Thus, the Vangioni-Flam et al.\ net uncertainties
are {\em lower limits} to the true uncertainties;
this fact is borne out 
in Figures \ref{fig:bbn-hi/lo}a and
\ref{fig:bbn-hi/lo}b, where the errors are equal to
or larger than those of
Vangioni-Flam et al.

\subsection{Minimal Uncertainties:  $\Delta \chi^2 = 1$}

\begin{figure}[htb]
\begin{center}
\vskip -0.5in
\mbox{ \epsfig{file=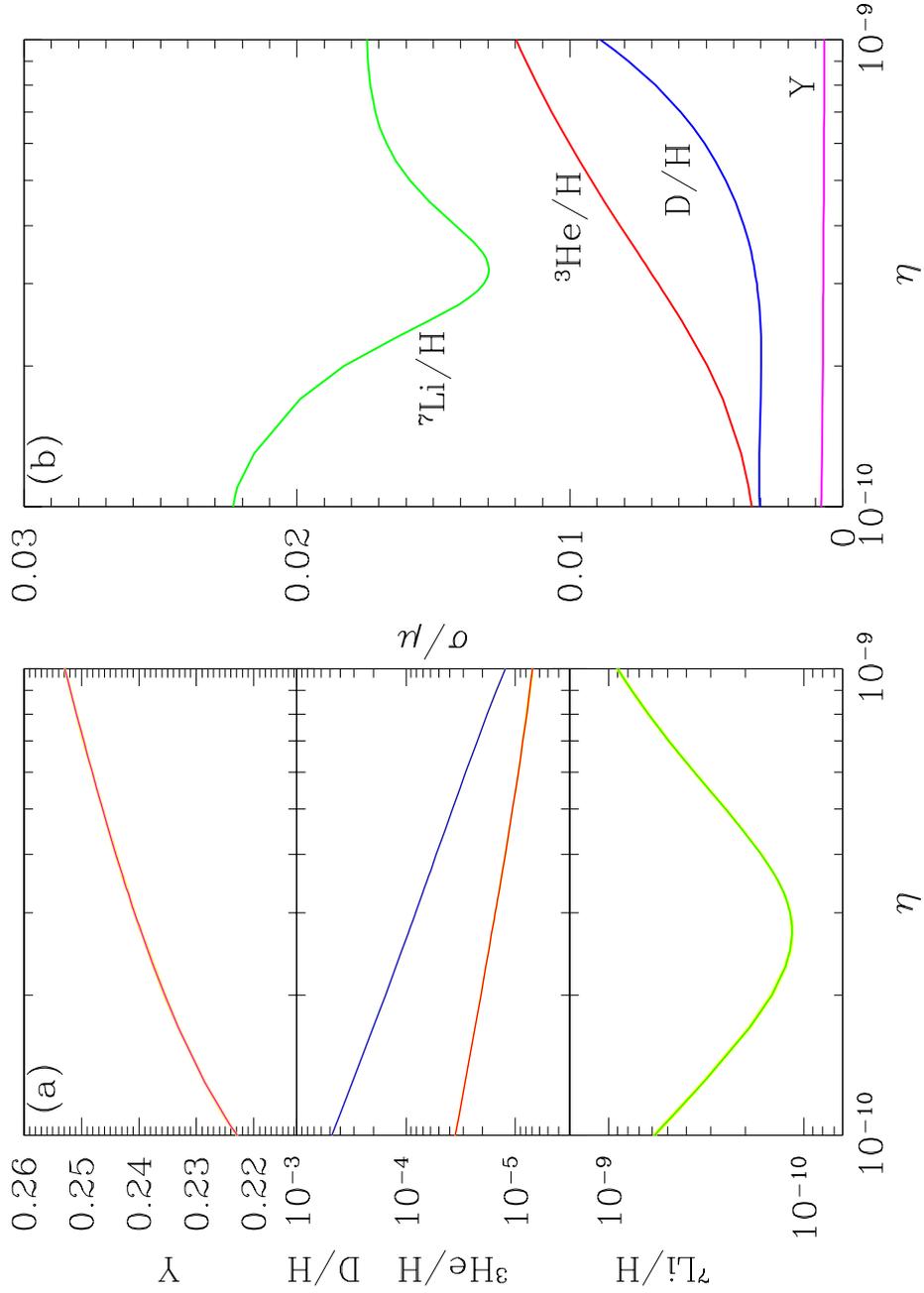,width=15cm}}
\end{center}
\vskip -0.5in
\caption[.]{\label{fig:bbn-minerr}
As in Figure \ref{fig:bbn-hi/lo}, with
errors for 
the $\Delta \chi^2 = 1$, ``minimal uncertainty'' estimates.
Note the change in vertical scale between
panel (b) and Figure \ref{fig:bbn-hi/lo}b.
}
\end{figure}

We now turn to the case of the uncertainties
given in the 
$\Delta \chi^2 = 1$ analysis of
\S \ref{sect:delta-chi2}.
Results appear in Figure \ref{fig:bbn-minerr},
where we see that the errors 
are now very small.
This is not a surprise, given that these
errors represent a lower limit to the true uncertainties.
In this limit, the nuclear input errors are small
enough to be negligible in comparison to the 
uncertainties one can expect in the observed abundances.

The errors presented here are derived by assuming that all
of the cross section data points may be combined without
concern for systematic errors between the experiments.
As this is not likely to be the case, the
true BBN theoretical error budget is considerably larger,
however.  We now turn to the ``sample variance'' error
estimator, which allows for these effects.

\subsection{Sample Variance}
\label{sect:sampvar}

\begin{figure}[htb]
\begin{center}
\vskip -0.5in
\mbox{ \epsfig{file=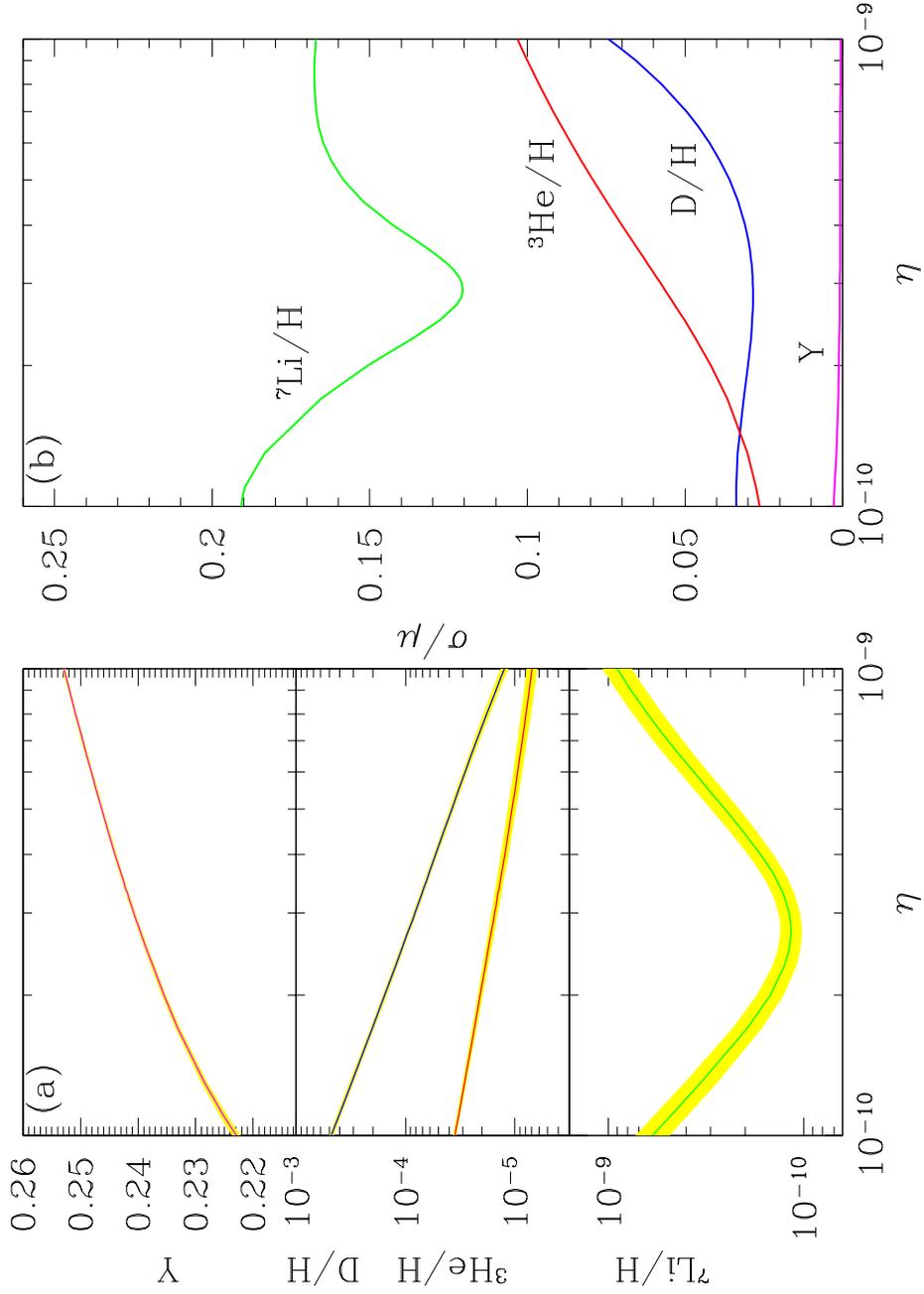,width=15cm}}
\end{center}
\vskip -0.5in
\caption[.]{\label{fig:bbn-sampvar}
As in Figure \ref{fig:bbn-hi/lo}, with
sample variance error estimates.

}
\end{figure}

The BBN predictions based on the sample variance 
errors appear in Figure \ref{fig:bbn-sampvar}.
The smallest errors are of course for \he4, where
the uncertainties are of order $\sim 2\%$, or $\Delta Y = 0.0005$. 
We note that the neutron lifetime is
now sufficiently well-known that the 
prediction for \he4 are now dominated instead
by the nuclear errors.
The D/H errors are also of the same order as the statistical errors
in the (high-redshift) data,
though systematics are likely to dominate the observational
uncertainties, as discussed in the next section.

The largest errors are for \li7, 
where the $\sim 10-18\%$ uncertainty is of the same order
of the {\em statistical} error in the observations.
Again, systematic errors are more important
for the observations.
Even so, it is particularly important to reduce the error
in Li {\em regardless} of the uncertainty in the observations.
As noted by, e.g., Ryan et al.\ \pcite{rbofn},
since the interesting range in $\eta$ places
the Li($\eta$) curve near its minimum, 
it is therefore slowly varying.
Even if there are {\em zero} observational errors,
the shallowness of the $\li7-\eta$ curve means that the
theoretical uncertainty in the Li prediction leads to a large
error in the allowed range for $\eta$ (below,
\S \ref{sect:like}).

We now compare the errors derived from
the sample variance with the other
estimators we have used. 
Referring to the panels (b)
of Figures \ref{fig:bbn-sampvar},
\ref{fig:bbn-hi/lo}, and \ref{fig:bbn-minerr}, 
we see that
the sample variance errors are smaller than those using
SKM, or NACRE's high/low ranges, but larger than 
the minimal errors, as expected.
Relative to SKM, the 
Li errors are reduced on the
low-$\eta$ side (i.e., the \li7 regime)
by the inclusion of the precision $t(\alpha,\gamma)\li7$,
data of Brune, Kavanagh, \& Rolfs \pcite{brune}.
On the high-$\eta$ side, the reduced Li errors 
follow from our smaller errors in $\he3(\alpha,\gamma)\be7$.
The \he3 and D errors are considerably lower
(by roughly a factor $\sim 2$),
due to the smaller errors in $p(n,\gamma)d$,
$d(d,n)\he3$, and $d(d,p)t$.

Of the error estimation methods we have used, 
we feel the sample variance method provides a good and fair estimate
of BBN uncertainties.  This procedure has the virtue of allowing
for systematic uncertainties due to reaction normalization
errors.  It also  has the virtue of simplicity of definition
and implementation, and one can judge by eye the goodness
of the fit.

\begin{figure}[htb]
\begin{center}
\mbox{ \epsfig{file=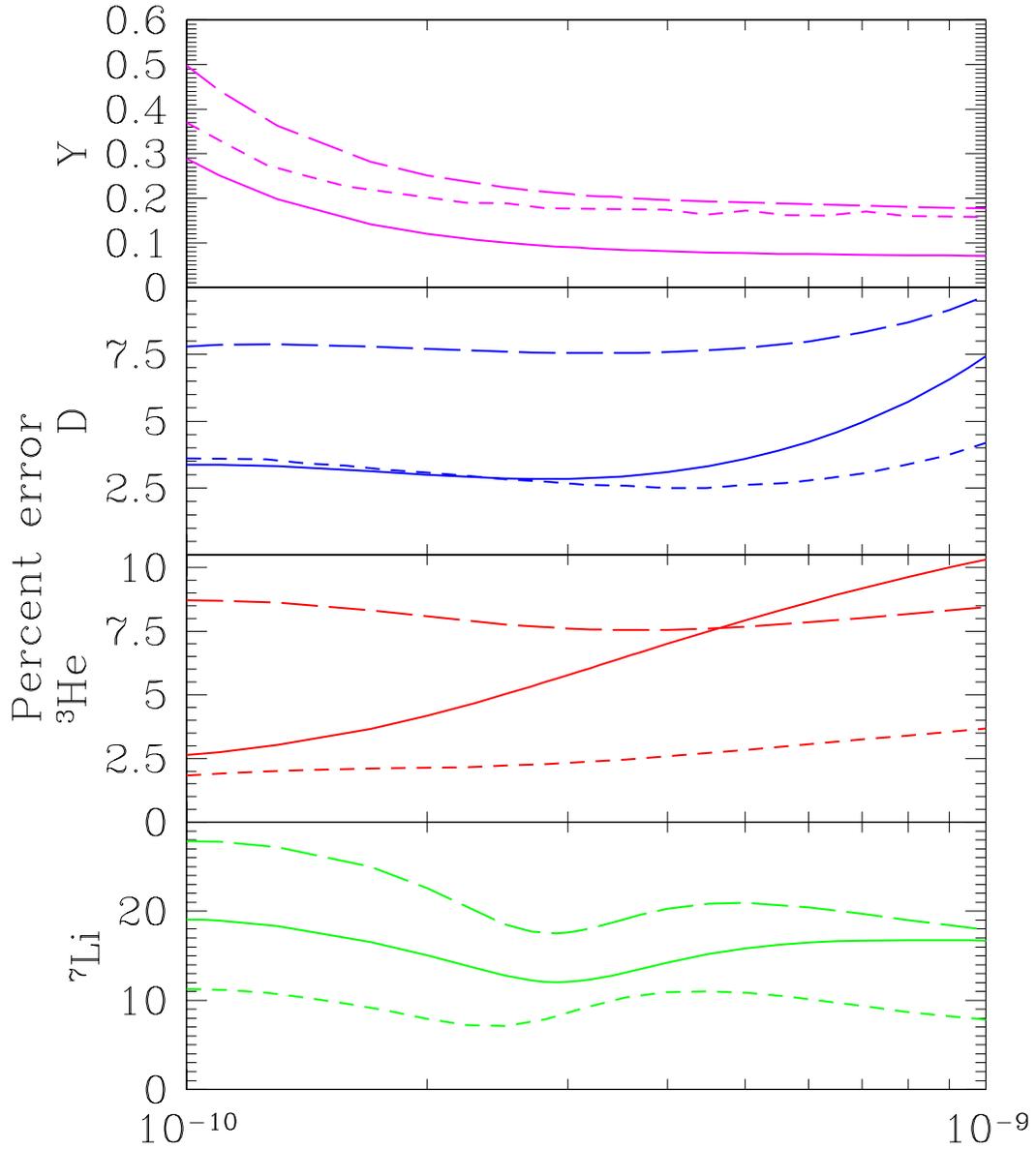,width=15cm}}
\end{center}
\vskip -1in
\caption[.]{\label{fig:errcomp}
A comparison of BBN fractional errors by different
groups.  Solid curves show our sample variance errors for the four
light elements, short dashed curves are the errors from Nollett \&
Burles
\cite{nb}, and long dashed curves are the errors from SKM.
We see that our errors generally fall between these two cases.
}
\end{figure}

The sample variance errors are compared with those
of Nollett \& Burles \pcite{nb} and
SKM in Figure \ref{fig:errcomp}.
The differences among the fractional errors plotted
stem mostly from differences in the errors $\sigma_i$ themselves,
except for the case of \li7, where the differences in
central values $\mu_i$ (ours are lower than those of SKM and NB)
also contribute to changes in $\sigma/\mu$.
We see in Figure \ref{fig:errcomp}
that our relative errors generally lie between
those of SKM and Nollett \& Burles.
This is as expected, since the NACRE thermonuclear rates, normalized or
not, follow similar trends to those rates used by SKM and Nollett \&
Burles.  Also, our sample variance usually lies between those of SKM and
Nollett \& Burles' errors.
For each element, different reactions dominate the
errors at different regimes in $\eta$, and thus
the variations of the shapes of the curves in each
panel of Figure \ref{fig:errcomp} reflect the differences
in the underlying reaction errors.
As seen in Figure \ref{fig:err-compare} and Table \ref{tab:errors},
both the SKM and Nollett \& Burles errors are generally of comparable
magnitudes across most reactions; ours have somewhat larger
variations in magnitudes, and thus our curves have
a somewhat different $\eta$ dependence than those
of these compilations.

For example,
in comparison with Nollett \& Burles,
we find that the D/H fractional errors are in very good agreement
for $\eta_{10} \la 3$.  At higher $\eta$, our errors then grow larger,
about 30\% larger at $\eta_{10} = 5$, to 80\% larger
at $\eta_{10} = 10$; this difference is mostly due to
the increasing sensitivity of deuterium to the $d(p,\gamma)\he3$ rate,
for which our adopted error is larger.
The trend for \he3 shows a more pronounced rise
at higher
$\eta$, due to the increasing importance of
both the $d(p,\gamma)\he3$ and $\he3(d,p)\he4$ rates,
where again we have used a larger uncertainty.

For \li7, our errors are consistently higher, by
a factor that varies from 70\% to 100\%;
part of this difference arises from the somewhat higher central
values Nollett \& Burles
have at higher $\eta$ (and thus a smaller $\sigma/\mu$).
The rest of the differences trace back to the systematically
tighter Nollett \& Burles error bands,
as discussed above.  The difference does not stem from
a single reaction, but rather the cumulative effect of the
lower Nollett \& Burles errors for several reactions:
$d(p,\gamma)\he3$, as well as the mass-7 producing and destroying
reactions.  In the case of $Y$, our errors are smaller than
those of the other groups; this arises because we have
adopted the new
errors in the neutron mean life which are
smaller than those used
by Nollett \& Burles and others in the past.

\section{Comparison with Observation}
\label{sect:like}

The procedure for comparing the light element abundances as predicted by
BBN theory with the light elements as observed astrophysically
is well-described elsewhere,
e.g., (Olive, Steigman, \& Walker \pcite{osw00}). 
Here, we simply summarize the
needed observational inputs and statistical techniques.

\subsection{Observational Inputs}

Data for \he4 is available from about 70 extragalactic \ion{H}{2}
regions (Pagel \etal 1992; Skillman \& Kennicutt 1993;
Skillman \etal 1994; Izotov, Thuan,   \& Lipovetsky 1994,1997;
Izotov \& Thuan 1998b) and has been recently compiled in 
Olive, Steigman, \& Skillman \pcite{oss} and
Fields \& Olive \pcite{fdo2}.  Unfortunately, the determination of
the \he4 abundance in these systems is not particularly straightforward.
To convert the observed He emission line strengths into abundances
requires knowledge of several physical parameters describing the \ion{H}{2}
system, such as the temperature, electron density, optical depth,
and degree of underlying stellar absorption.

The older data of Pagel \etal (1992), Skillman \& Kennicutt (1993),
and Skillman \etal (1994), used \ion{S}{2} data to fix the
electron density from which was derived a relatively
low \he4 abundance.  Higher \he4 abundances were derived by
Izotov, Thuan,  \& Lipovetsky (1994,1997) and 
Izotov \& Thuan (1998b), where the electron density was determined
self consistently from five distinct He emission lines. 
In addition, it was pointed out in Izotov \& Thuan (1998a) that one
of lowest metallicity regions, I Zw 18 NW, is plagued by underlying
stellar absorption.  The compilation by Izotov \& Thuan (1998b) 
of 45 extragalactic \ion{H}{2} regions yielded a primordial \he4 abundance
(by performing a regression on the data versus the O/H abundance)
of $Y_p = 0.244 \pm 0.002$. As the compilation of Izotov \& Thuan
(1998b) also presented results based on \ion{S}{2} densities, it was
possible to combine all of the data in a systematic way. The result
of Fields \& Olive \pcite{fdo2} which included the data of 
Pagel \etal (1992), Skillman \& Kennicutt (1993),
and Skillman \etal (1994) and Izotov \& Thuan (1998b) (but without
the possibly erroneous I Zw 18 NW), yielded the somewhat lower
result of $Y_p = 0.238 \pm 0.002$. Note that the data of Izotov \&
Thuan (1998b) alone based on \ion{S}{2} densities give
$0.239 \pm 0.002$.  One should
also note that a recent determination by
Peimbert, Peimbert, \& Ruiz (2000) of the \he4 abundance in a
single object (the SMC) also using a self consistent method gives
a primordial abundance of 0.234 $\pm$ 0.003 (actually, they observe
$Y = 0.240 \pm 0.002$ at the relatively high value of [O/H] = -0.8,
where [O/H] refers to the log of the Oxygen abundance relative to
the solar value).

Recently, a detailed examination of the systematic uncertainties in
the \he4 abundance determination was made by Olive \& Skillman
(2000). There a Monte Carlo simulation of synthetic data showed that
the He abundance determinations using a $\chi^2$ minimization in the
self-consistent method typically under-estimated the true errors by
about a factor of 2.  The reason for the enhanced error
determinations is a degeneracy among the physical parameters 
(electron density, optical depth, and  underlying stellar
absorption) which yield equivalent $\chi^2$ results.  Indeed, it was
found that there are competing biases in the data.  On the one hand,
the presence of underlying stellar absorption (expected to be
present to some level in most systems) leads to an underestimate of
the He abundance. While on the other hand, the minimization
solutions tend to find solutions with erroneously low densities
thus minimizing the collisional corrections and over-estimating the 
He abundance.

With these cautions in mind, we will never-the-less adopt 
the Fields \& Olive \pcite{fdo2}
value of 
\beq
Y_p = 0.238 \pm 0.002 \pm 0.005
\eeq
where the first error is statistical and the second systematic.
Clearly, the  dominant error source is systematic,
and while we have given our best estimate for its value,
this number may yet turn out to be too small until a
complete reanalysis of the  existing  or new data is possible. 
For the full error in $Y_p$, we will use the quadrature sum of the
two components.

As in the case of \he4, there is a considerable body of data on \li7.
There are well over 100 hot halo dwarf stars with \li7 observations.
The discovery of a plateau in the Li abundance for hot and metal
poor dwarfs by Spite \& Spite (1982), is generally recognized as the
primordial value. Recent high precision studies of Li abundances in
halo stars have been made by Bonifacio
\& Molaro (1997) and  Ryan, Norris, \& Beers \pcite{rnb} which
confirm the plateau. The latter in fact identified a small but
significant Li-Fe trend in the low metallicity regime.
Ryan et al.\ \cite{rbofn} showed that this trend is
indeed {\em expected} given the presence of cosmic ray 
interactions in the early Galaxy.
These provide a source of Galactic Li production on top of
a cosmological value (very analogous to the 
derivation of $Y_p$ from a regression of  $Y-Z$ data and the $dY/dZ$ slope).
Ryan et al.\ inferred a primordial Li abundance of 
\beq
\label{eq:Li_p}
\li7/{\rm H} = (1.23^{+0.68}_{-0.32}) \times 10^{-10}
\eeq
Note that correcting for Galactic production 
{\em lowers} \li7/H compared to taking the mean value
over a range of metallicity.  
For our adopted primordial \li7, we assume
a gaussian distribution with central value equal to that of 
eq.\ \pref{eq:Li_p}, and a standard deviation which we conservatively take
to be the larger of the asymmetric errors.

We note that in contrast to the downward correction
due to post big bang production of Li, there is a potential for an upward
correction due to depletion.  While most studies of possible depletion
factors are small, they may not be negligible.  Both 
Vauclair \& Charbonnel \pcite{vc}
and Pinsonneault \etal (1998) argue for an upward correction of 0.2
dex.  We note however, that the data do not show any dispersion
(beyond that expected by observational uncertainty) and both 
 Bonifacio
\& Molaro (1997) and  Ryan, Norris, \& Beers \pcite{rnb} have argued
strongly against depletion.  We further point out that the
observation of the fragile isotope, \li6, (Smith, Lambert \& Nissen
1992, 1998; Hobbs and Thorburn 1994, 1997, Cayrel 
\etal 1999; Nissen \etal 2000) also puts strong
constraints on the degree of depletion (Steigman \etal 1993;
Lemoine \etal 1997; Fields \& Olive 1999; Vangioni-Flam \etal
1999).  

The observational status of primordial D is 
complicated but also promising.
Deuterium has been detected in several
high-redshift quasar absorption line systems.
It is expected that these systems still retain
their original,  primordial deuterium, unaffected by any significant
stellar nucleosynthesis.
At present, however, there are four good determinations of D/H in
absorption systems with considerable scatter.  Two
good measurements  (Burles \& Tytler \cite{bt98a,bt98b})
with D/H determinations of 
D/H = $(3.3 \pm 0.3) \times 10^{-5}$ and $(4.0 \pm 0.7) \times
10^{-5}$, have recently
been supplemented by 
the new results of 
O'Meara \etal\ \pcite{omera} with D/H = $(2.5 \pm 0.2) \times
10^{-5}$. These authors measured D/H in 
a Lyman limit system which has a
neutral hydrogen column density 
a factor of 30--100 higher than the previous
D/H systems, and in which most of the H is neutral
rather than ionized.  As a result, 
O'Meara \etal\ \pcite{omera}
were able to detect D in five transitions.
The inferred D/H value is lower than the previous
absorption systems, and the three
systems combined give 
\beq
{\rm D/H} = (3.0 \pm 0.4) \times 10^{-5}
\label{lowd}
\eeq
O'Meara \etal\ \pcite{omera} note, however, that $\chi^2_\nu = 7.1$ for 
the combined 3 D/H measurements (i.e., $\nu = 2$), 
and interpret this as a likely indication that the
errors have been underestimated.

Significantly more discrepant is the higher determination of
D/H, $(2.0 \pm 0.5) \times 10^{-4}$, as seen in one low-redshift
system  
(Webb et al.\ \cite{webb}; Tytler et al.\ \cite{tytler}).
Clearly there are systematic uncertainties which are yet to be
sorted out. Indeed, Levshakov, Kegel \& Takahara (1998)
have used the  data in the high redshift system of Burles \& Tytler
\cite{bt98a} but  with a different model
for the velocity distribution of the absorbing gas,  and derived a
95\% confidence range 3.5 $\times 10^{-5} \leq$~D/H~$\leq 5.2 
\times 10^{-5}$.  Levshakov, 
Tytler \& Burles \cite{lev} have also applied this 
model to a reanalysis of the  system 
of Burles \& Tytler
\cite{bt98b}, finding a a 68\% confidence range of  
D/H $\simeq (3.5 - 5.0) \times 10^{-5}$.
 Because of these complications, we will examine
the impact of low and high D on BBN.

\subsection{Likelihood Analysis}

To quantitatively compare the observed abundance data with theory, 
we use a likelihood analysis.
The formalism is described elsewhere
(Fields \& Olive \cite{fdo1}; 
Fields, Olive, Kainulainen, \& Thomas \cite{fkot}),
but the basic idea is that the Monte Carlo
results (means, variances, correlation coefficients) allow
one to compute the theory likelihoods for the elements
as a function of $\eta$.  The convolution of these with
the observed abundance distributions gives the
overall likelihood as function of $\eta$; these can
determine the goodness of fit of the model and
allow for parameter estimation.

\begin{figure}[htb]
\begin{center}
\mbox{ \epsfig{file=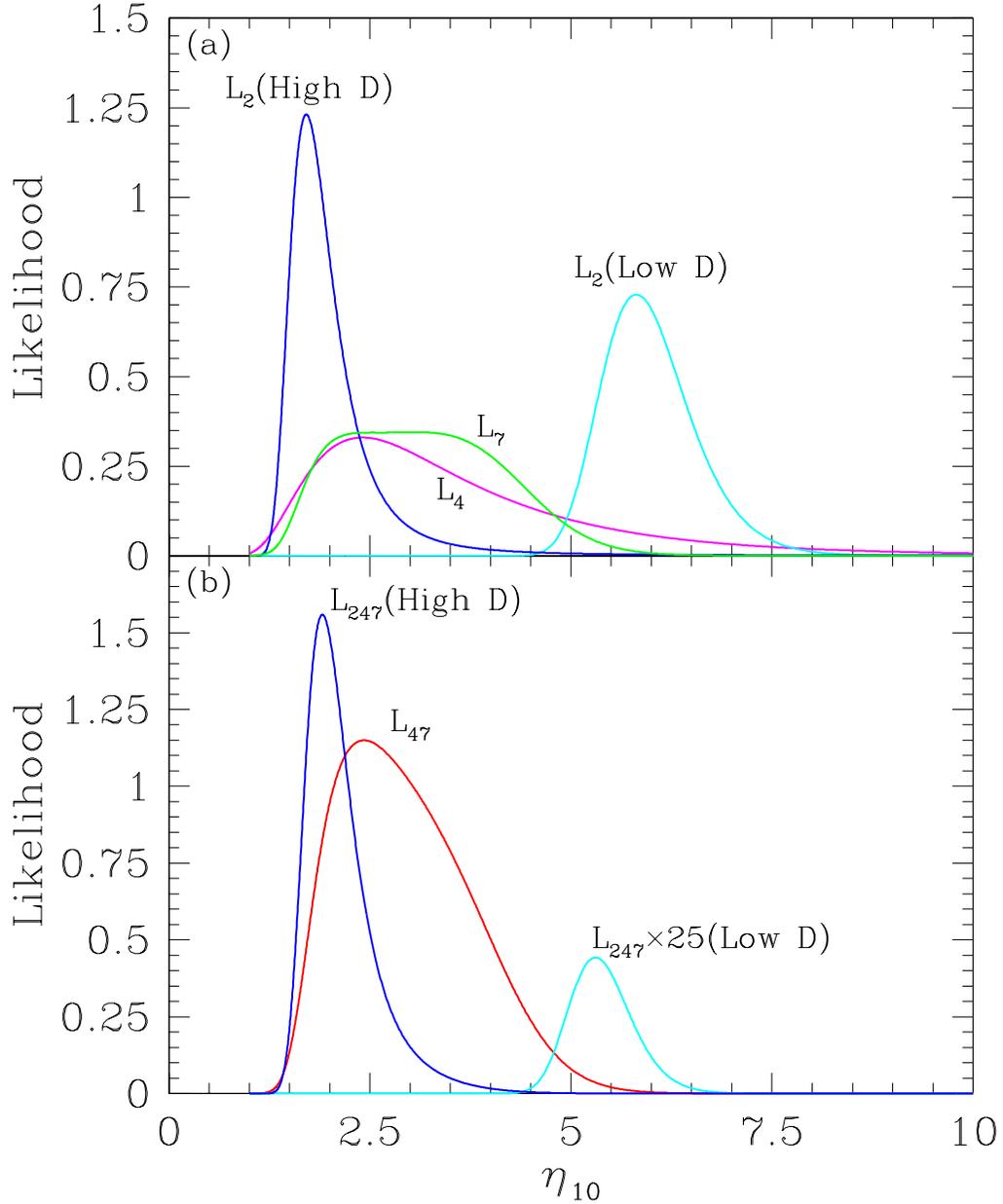,width=15cm}}
\end{center}
\vskip -1in
\caption[.]{\label{fig:like-sampvar}
Likelihoods for the sample variance errors.
a)
Curves for \he4 and \li7, as well 
as high and low D individually.
Plotted are the likelihoods for combined theory and observation
distribution. 
The small amplitude for the low D case indicates
a poorer fit.
b)
Total 2- and 3-element combined likelihoods.
}
\end{figure}

Figure \ref{fig:like-sampvar}
shows the likelihoods derived using
the sample variance errors.
In panel (a), we see the theory-observation combined
distributions for individual elements.
The peaked nature of the \he4 and D curves
stems from the monotonic nature of the $\eta$ dependence of
these abundances.  The flatness of the \li7 likelihood
arises because the observed \li7 abundance we have adopted
lies at the minimum of the lithium trough, so that the
usual two-peaked structure has merged into a single
broad feature.
As has noted in Fields \etal\ \cite{fkot} and elsewhere,
the \he4 and \li7 likelihoods are in excellent agreement
with each other and with
the high D distribution.  This is reflected
in the combined likelihood distribution (panel b).
Here, $L_{47}$ represents the combined likelihood distribution using
only \he4 and \li7, while $L_{247}$ is the combined likelihood
distribution using D, \he4 and \li7. A value of order
unity is an indication that the agreement is good.

On the other hand, the
low D distribution agrees marginally at best with
the \he4 and \li7 data.
This discrepancy has been widely noted,
and if real, suggests that either one or more
of the observed abundances is incorrect, that the systematic uncertainties
have been grossly underestimated, or perhaps that new physics is at play.
One should recall, though, that
the D observations could as well suffer from
systematic errors that are not reflected in 
numbers we have adopted.  Indeed, systematic uncertainties are not
included in the error budget of eq. (\ref{lowd}).  Since D/H is a strong
function of
$\eta$, the D likelihood distribution is very sensitive to changes in the D
data; for example, a modest shift upward in the D central value, or a
larger D error, could lead to  agreement with \he4 and \li7.
For example, even if D/H is only as high as $\sim 5 \times 10^{-5}$, as
deemed possible in the mesoturbulent models of 
Levshakov, Kegel \& Takahara (1998) and Levshakov, 
Tytler \& Burles \cite{lev}, the peak of the likelihood function shifts
down to $\eta_{10} \approx 4$ and would indeed be in good agreement with
the other light elements.  As is also well known, the high D/H
distribution is in excellent agreement as seen in Fig.
\ref{fig:like-sampvar}.  Clearly, a determination of the true, primordial
D  abundance--and an accurate assessment of its errors---is of paramount
importance.

\begin{figure}[htb]
\begin{center}
\mbox{ \epsfig{file=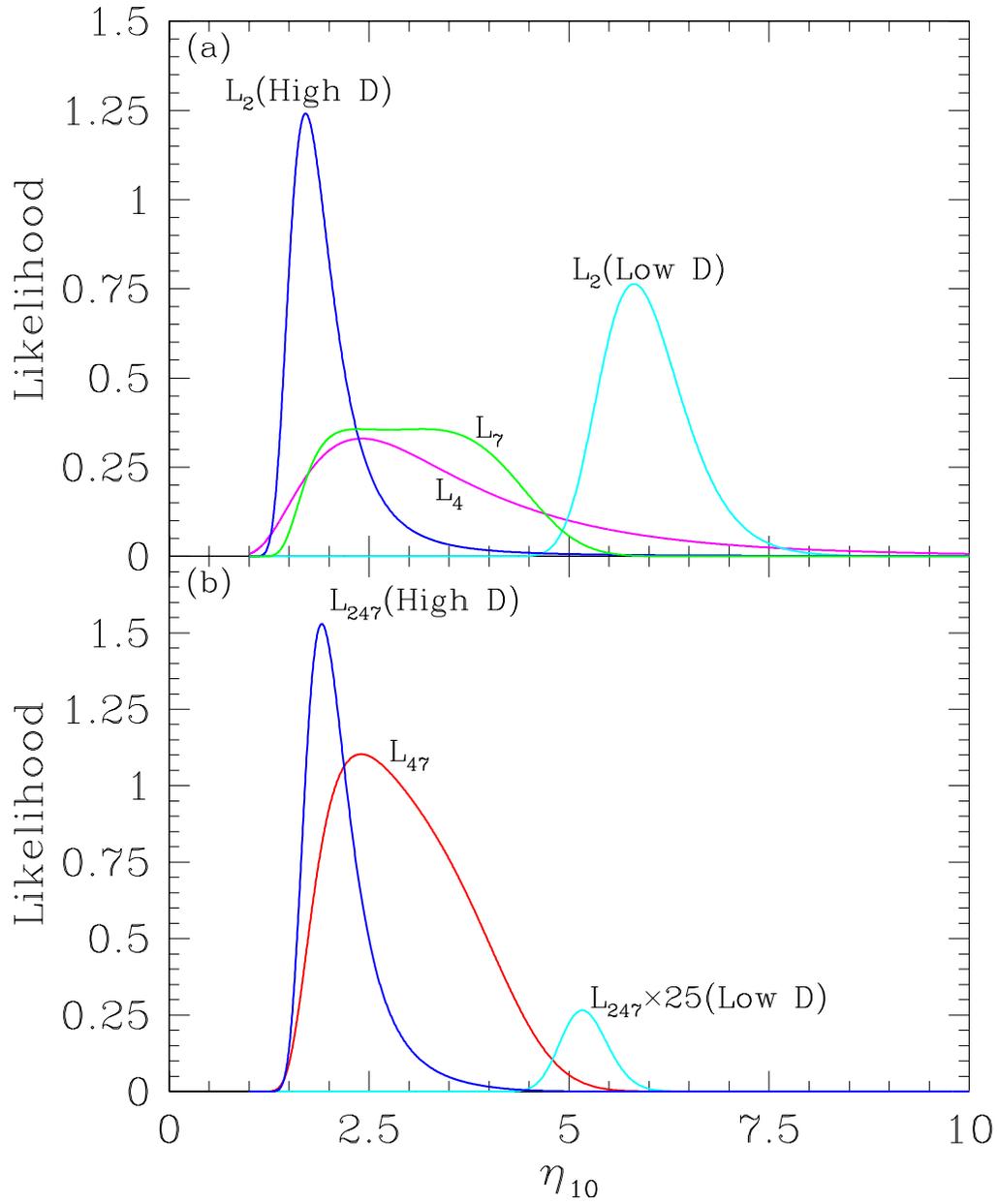,width=15cm}}
\end{center}
\vskip -1in
\caption[.]{\label{fig:like-tiny}
Likelihoods as in Figure
\ref{fig:like-sampvar}
for the minimal errors.
}
\end{figure}

Had we used the high/low errors
of the NACRE rates
we would have obtained very similar likelihood distributions and
predictions for $\eta$.  Similarly, the use of the unrenormalized rates 
would not have made a big difference here either.  In the extreme case
that we use our minimal uncertainties based on $\Delta \chi^2 = 1$, we
would of course have tighter predictions as shown in
Figure \ref{fig:like-tiny}.
It is interesting to note that even in this case, the
likelihood distributions for \he4 and \li7 are broad due to
residual systematic uncertainties in the observations.
Therefore, for any significant improvement in the BBN
predictions, the systematic uncertainties will need to be
resolved.

\subsection{Constraints on Cosmology:  The Baryon Density and Light Neutrinos}
\label{sect:eta}

Using the results of the likelihood analyses in
the previous section, we can derive limits on $\eta$.
We will focus on the sample variance case;
as we have remarked, the likelihoods for the high/low
case are only slightly different, and we
have verified that they give very similar results
for $\eta$.

For the sample variance error, 
we derive a most-likely value of $\eta$, $\hat{\eta}$, and a
95\% CL range.
For the \he4-\li7 combined likelihood,
we find (to two significant digits)
\beqar
\label{eq:eta47}
\hat{\eta}_{10} = 2.4& \Rightarrow  &\hat{\Omega}_{\rm B} h^2  =
0.0089    \\ 1.7  \le {\eta_{10}} \le 4.7
  & \Rightarrow & 0.006 \le \Omega_{\rm B} h^2 \le 0.017 \ \ \
\mbox{(95\% CL) \ .}
\eeqar
where $\hat{\eta}$ is the value at maximum likelihood.
We use the conversion $\eta_{10} = 274 \omegab h^2$.
For high D, the result is very similar, owing to the
close overlap of the individual element likelihoods; we have
\beqar
\label{eq:eta247h}
\hat{\eta}_{10} = 1.9& \Rightarrow  &\hat{\Omega}_{\rm B} h^2  =
0.007     \\ 1.6  \le {\eta_{10}} \le 3.3
  & \Rightarrow & 0.006 \le \Omega_{\rm B} h^2 \le 0.012 \ \ \
\mbox{(95\% CL) \ .}
\eeqar
For low D, the goodness of fit is worse, as we have noted,
and we have 
\beqar
\label{eq:eta247l}
\hat{\eta}_{10} = 5.3& \Rightarrow  &\hat{\Omega}_{\rm B} h^2  =
0.019    
\\ 4.7  \le {\eta_{10}} \le 6.3  
  & \Rightarrow & 0.017 \le \Omega_{\rm B} h^2 \le 0.023 \ \ \
\mbox{(95\% CL) \ .}
\eeqar

We have of course used the standard model in the above determinations of 
$\eta$.  In particular, we have assumed $N_\nu = 3$. As was shown in 
previous analyses (Olive \& Thomas 1999, Lisi, Sarkar, \& Villante 1999),
the combination of \he4 and \li7 (as well as high D/H) is quite compatible
with the standard model value, and we do not expect that result to change
here.  Similarly, the result including low D/H is at best compatible with
$N_\nu = 3$ at the 2 $\sigma$ level.  A complete two-dimensional
likelihood analysis based on the NACRE compilation is beyond the scope of
the present work.

\section{Constraints from Cosmology:  The Microwave Background}
\label{sect:cmb}

BBN has long provided the best estimate
of $\eta$, and thus the cosmic baryon content
(i.e., $\Omega_B h^2$).  However,
measurements of the anisotropy spectrum of the CMB
can constrain many key cosmological parameters,
including $\eta$ (see e.g., White, Scott, \& Silk
\cite{wss}).   Thus, the agreement between the two
estimates of cosmic baryons provides a fundamental test
of cosmology (Schramm \& Turner \cite{st}).  

Recent anisotropy data already allows for an initial 
investigation of the range in $\eta$
favored by CMB.
The acoustic peaks in the CMB at small angular
scales are sensitive to the baryon density as well
as to other parameters, and the most reliable estimates
of $\eta$ come from fitting over a range of angular
scales (including multiple peaks and valleys).
The BOOMERanG-98
(de Bernardis \etal\ \cite{boom})
and MAXIMA-1 (Hanany \etal\ \cite{maxima};
Balbi et al.\ \cite{balbi}) experiments
provide the first data well-suited for this effort, 
as they span the entire first acoustic peak and 
the region where the second peak is expected.

For these experiments, the
inferred value of $\eta$ is sensitive mostly to 
ratio of first to second peaks.
The second peak appears weakly, if at all, in 
the BOOMERanG-98 data, and somewhat more strongly
in the MAXIMA-1 data.
The two groups 
combined their data (and those of COBE), and made
estimates for a set of cosmological parameters 
(Jaffe \etal\ \cite{combo}).
They estimate $\Omega_B h^2 = 0.032^{+0.005}_{-0.004}$ (68\% CL),
which corresponds to
\beq
\label{eq:eta-cmb}
\eta_{\rm CMB} = (8.8^{+1.4}_{-1.1}) \times 10^{-10} \ \ \mbox{(68\% CL)}
\eeq
Similar analyses by other authors 
give similar results\footnote{
In the cases where nucleosynthesis information was
not assumed as a prior.}
(e.g., Tegmark \& Zaldarriaga \cite{tz};
Tegmark, Zaldarriaga, \& Hamilton \cite{tzh}).

The CMB preferred range in $\eta$ (eq.\ \ref{eq:eta-cmb})
is about $3 \; \sigma$ away from the 
BBN results derived from \he4 and \li7 
(Figure \ref{fig:like-sampvar}b and eq.\ \ref{eq:eta47}).
Including low D/H only slightly reduces the difference, 
to two $\sigma$.  
Indeed, the CMB favored range based on the BOOMERanG-98 and MAXIMA-1 data
does not agree with {\em any} of the three light nuclides.
For example, the CMB data demand 
that ${\rm D/H} = 1.6 \times 10^{-5}$,
which is significantly below all high-redshift
determination of deuterium.
Moreover, the CMB-preferred D/H lies
just at the present-day local interstellar value,
${\rm D/H} = (1.5 \pm 0.1) \times 10^{-5}$ 
(e.g., Linsky \etal\ \cite{linsky}; Sahu \etal\ \cite{sahu}).
This would allow for almost no processing of D/H over the
history of the Galaxy.  This is very unlikely even if
primordial infall played a big role in the chemical
evolution of our Galaxy. The CMB data also require
$Y_p = 0.252$ and
${\rm \li7/H} =
\ee{7.4}{10}$. For \he4, this is somewhat high, but given
the magnitude of the systematic uncertainties, it would be
difficult to exclude this value outright.  On the other
hand, the \li7 value would require a factor of $\sim$ 6 in
depletion.  This is probably a factor of $\sim$ 3 too high
relative to the most optimistic models of stellar depletion. 

If the discrepancy is real, it may point to 
new physics occurring sometime prior to recombination.
Kaplinghat \& Turner \cite{kt} pointed out that 
entropy production 
after BBN but before recombination 
could account for the difference without
requiring new physics at the BBN epoch.
On the other hand, new BBN physics could also
lead to a higher baryon density for the same
abundance constraints.
In particular, there has been considerable 
attention recently to the case
of BBN with a large neutrino chemical potential
(Esposito, Mangano, Melchiorri, Miele, \& Pisanti \cite{emmmp};
Lesgourgues \& Peloso \cite{lp};
Orito, Kajino, Mathews, \& Boyd \cite{okmb};
Esposito, Mangano, Miele, \& Pisanti \cite{emmp00b}).
Finally, Kneller, Scherrer, Steigman, \& Walker \cite{kssw},
show that while the BBN-CMB discrepancy is robust given the BOOMERanG-98
and MAXIMA-1 data, the nonstandard parameters
depend sensitively on CMB priors.

However, before we abandon the standard model,
we note also CMB results of the two experiments
independently have very different second
peak points, leading to different (but consistent) baryon contents.
Furthermore, the recent data
from the Cosmic Background Imager (CBI; Padin et al.\ \cite{cbi}) implies
a very different result. CBI is sensitive to the higher order acoustic
peaks ($\ell = 400 -1500$), and their first results
favor $\Omega_B h^2 = 0.009$, or $\eta_{10} = 2.46$.
While this is in sharp contrast with the results of BOOMERanG and MAXIMA, 
this value is {\em lower} than the $\eta$ range
suggested by the low D/H data.  In fact,
the CBI result is in excellent agreement with the $\eta$ range
favored by \he4 and \li7 (Figs \ref{fig:like-sampvar}).
Clearly, more and better CMB data are needed 
to clarify this situation.

Fortunately, high-quality CMB data will soon be at hand.
The MAP satellite (Wright \cite{map}) is scheduled to 
launch in 2001, and will return data about a year
later.  These precision measurements should give
$\Omega_B$ to better than 10\%.
The PLANCK Explorer, scheduled for launch in 2007, will reduce these
errors to better than 3\%.
As these data become available, the comparison of
the $\eta$ predictions with those of
BBN will become an acute test of cosmology.
Clearly, a strong discrepancy will be a surprise
and will demand an explanation.

\begin{figure}[htb]
\begin{center}
\mbox{ \epsfig{file=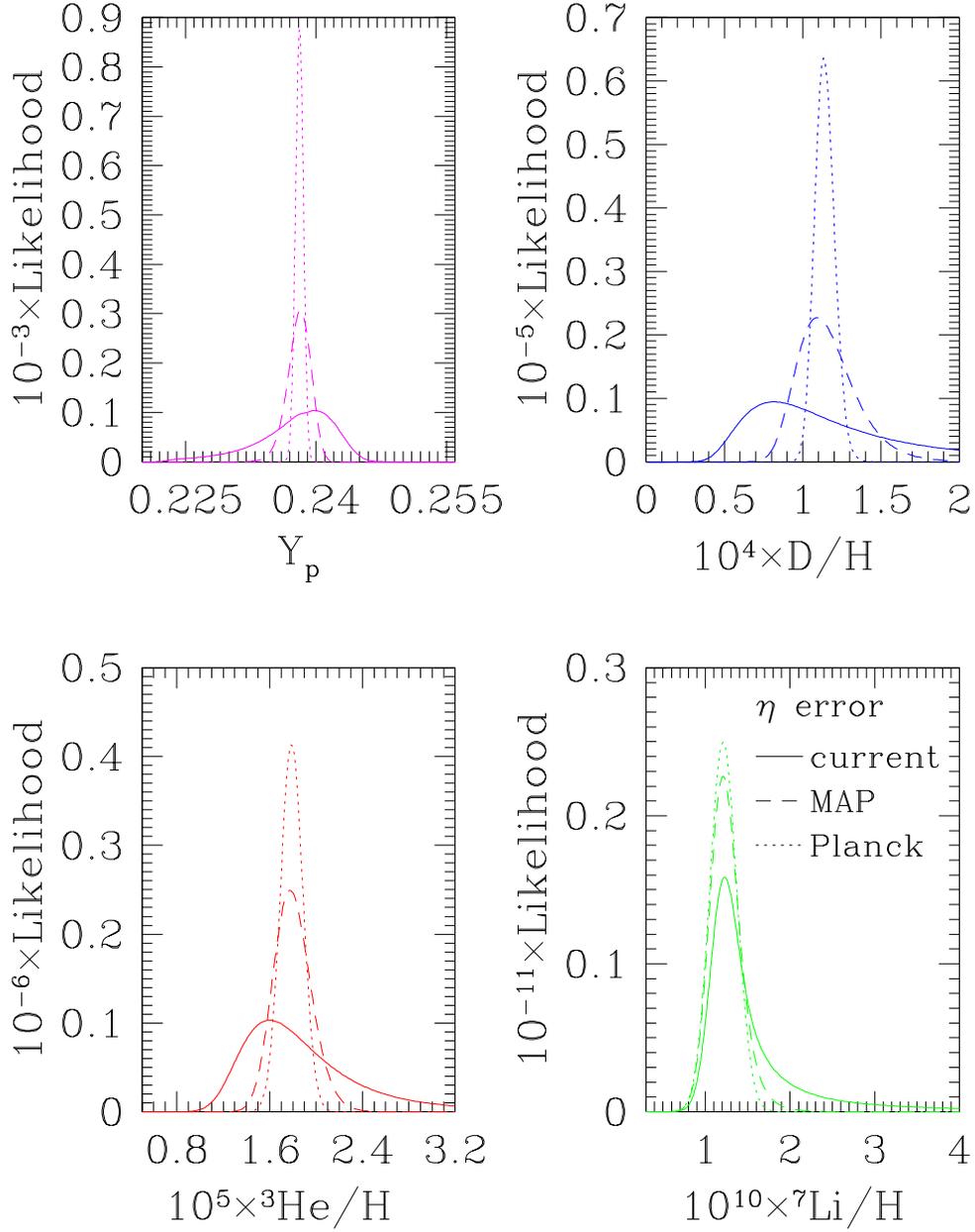,width=15cm}}
\end{center}
\vskip -1in
\caption[.]{\label{fig:cmb-future}
Light element abundances as might be predicted
by $\eta$ inputs from future CMB experiments.
We have assumed a central value of 
$\eta_{10} = \hat{\eta}_{10}$ (eq.\ \ref{eq:eta47}).
Solid curve:  10\% errors in $\eta$ (MAP);
Broken curve:  3\% errors in $\eta$ (PLANCK).
We see that predictive power of the CMB limits on $\eta$ will
become quite strong.
}
\end{figure}

On the other hand, it is possible that, with
precision results available for
both BBN and the CMB, the baryonic predictions will
be in agreement.  Then, as the CMB results improve,
it will become useful to use the CMB
range for $\eta$ as an {\em input} to the BBN analysis, and 
to use this, e.g., to give precision predictions for
the light element abundances.
An illustration of such a prediction is shown in
Figure \ref{fig:cmb-future}, which
shows the light element abundance predictions
assuming a constant central value of $\eta_{10} = \hat{\eta}_{10}$
(as in eq.\ \ref{eq:eta47}), but with
an error which goes from that of our current limits down to
10\% (MAP) and then 3\% (PLANCK).
We see that these experiments will be able to
determine the primordial abundances quite
accurately and thus will impact studies of
galactic, stellar, and cosmic-ray evolution.

A noteworthy feature of Figure \ref{fig:cmb-future}
is the slight shift in the peak values of the likelihoods as
the limits on $\eta$ become tighter.
Thus, the maximum likelihood estimator for
the abundance $y$ is slightly offset from
$y_{\rm BBN}(\hat{\eta})$ but converges to this
value as the CMB uncertainty in $\eta$ becomes small.
This comes about because the plotted likelihood
is given by the convolution
\beq
L(y) = \int d\eta \; L_{\rm CMB}(\eta) \; L_{\rm
BBN}(\eta;y)
\eeq 
The BBN theory likelihood
\beq
L_{\rm BBN} \propto \sigma_{\rm BBN}^{-1} \; 
  \exp[-(y-y_{\rm BBN})^2/2 \sigma_{\rm BBN}^2]
\eeq
depends on 
$\eta$ in a complicated way through the abundance
and error curves $y_{\rm BBN}(\eta)$ and $\sigma_{\rm BBN}(\eta)$.
One can show that the peak of $L(y)$ is shifted
due to the variation of the mean values and errors
with $\eta$.
One can also
show that the shift goes to zero as the CMB error
becomes small.

\section{Conclusions}
\label{sect:conclude}

We have examined the effect of the NACRE thermonuclear
reaction rates on BBN.
We verify that the central values of the
abundance predictions are very similar to those
used in previous studies (cf. Hata \etal 1996).
We note that the NACRE fits are in general good representations
of the shape of the data, but the normalizations
do not precisely minimize $\chi^2$.  We have thus
computed the renormalizations needed to minimize
$\chi^2$. Although small, the renormalizations do lead to changes
with the previous predictions, particularly in that 
the Li prediction at high $\eta$ is reduced,
and thus less Li depletion is required in this regime.

We have used the NACRE rates and database to
determine the theoretical uncertainty in BBN
via the standard Monte Carlo procedure.
To do this we have examined the effect of different estimates
of the nuclear reaction uncertainties.
NACRE's high/low limits
give BBN uncertainties
similar to those of SKM.
Using a simple sample variance procedure,
we arrive at errors which are smaller than these,
while being slightly larger than those of Nollett and Burles.
We have also computed lower limits to the errors on
the reaction rates, which lead to the minimal theoretical
errors on the light element abundances.

Error studies such as ours are increasingly important
as cosmology enters the 
precision era.
In particular, an independent prediction of $\eta$
is emerging from CMB anisotropy data.
At present, the CMB results are tantalizing, and possibly
discrepant with the BBN predictions, though the
CMB predictions for $\eta$ seem to be
uncertain at the moment.
At any rate, the CMB data will improve dramatically with the 
upcoming launch of MAP; this experiment, along with PLANCK,
will make a strong test of BBN and of cosmology.

In anticipation of this test,
further work is needed to find a realistic error budget for
both the theory and observation of light element abundances.
A key input on the theory side can come
from accurate nuclear experimental data and careful analyses
of the error budget in key reactions.
Specifically, we urge more data be taken for 
$p(n,\gamma)d$, in order to 
confirm theory, and put this reaction on a solid empirical basis.
Other key reactions are those with
internally discrepant reactions, 
$d(p,\gamma)\he3$ and
$\he3(d,p)\he4$,
and the lithium sources and sinks,
$t(\alpha,\gamma)\li7$,
$\he3(\alpha,\gamma)\be7$, and
$\li7(p,\alpha)\he4$.

\acknowledgments
We are grateful to the NACRE collaboration, 
and particularly Carmen Angulo and
Pierre Descouvrement, for their 
helpful responses to our questions about
the data and rate compilations.
We are indebted as well to Michael Smith for 
instructive discussions and sharing
reaction fitting functions.
We thank Sam Austin, Carl Brune, Michel Cass\'{e},
Alain Coc, Vijay Pandaripande, and Elisabeth
Vangioni-Flam
for instructive conversations.

\newpage

\end{document}